\documentclass[11pt]{article}

\usepackage{fancyhdr}

\usepackage{geometry}

\RequirePackage[numbers]{natbib}

\renewcommand{\Re}{\mathbb{R}}

\newcommand{\vat}{\mathbb{E}}

\newcommand{\bc}{\begin{center}}
	\newcommand{\ec}{\end{center}}

\newcommand{\bit}{\begin{itemize}}
	\newcommand{\eit}{\end{itemize}}

\newcommand{\be}{\begin{eqnarray*}}
	\newcommand{\ee}{\end{eqnarray*}}
\newcommand{\ben}{\begin{eqnarray}}
\newcommand{\een}{\end{eqnarray}}

\newcommand{\g}{\,\vert\,}

\newcommand{\D}{\mathcal{D}}

\newcommand{\N}{\mathcal{N}}
\newcommand{\W}{\mathcal{W}}

\newcommand{\pa}{\mathrm{pa}}
\newcommand{\fa}{\mathrm{fa}}

\newcommand{\bzero}{\bm{0}}

\newcommand{\bD}{\bm{D}}

\newcommand{\bI}{\bm{I}}

\newcommand{\bL}{\bm{L}}

\newcommand{\bS}{\bm{S}}

\newcommand{\bU}{\bm{U}}
\newcommand{\bX}{\bm{X}}

\newcommand{\bx}{\bm{x}}

\newcommand{\bSigma}{\bm{\Sigma}}

\newcommand{\bOmega}{\bm{\Omega}}

\newcommand{\bgamma}{\bm{\gamma}}

\newcommand{\btheta}{\bm{\theta}}
\newcommand{\bmu}{\bm{\mu}}

\usepackage{tikz}
\usetikzlibrary{graphs}
\usetikzlibrary{positioning}

\newcommand{\black}{\color{black}}

\usepackage{bm}
\usepackage{bbm}
\usepackage{float}
\usepackage{amssymb}
\usepackage{subfig}
\usepackage{enumerate}
\usepackage{longtable}
\usepackage{mathtools,cancel}
\usepackage{url}
\usepackage{multirow}
\usepackage{lscape}
\usepackage{authblk}

\usepackage{xr}

\begin{document}

\title{Bayesian graphical modeling for heterogeneous causal effects}
\author[1]{Federico Castelletti \thanks{federico.castelletti@unicatt.it}}
\author[2]{Guido Consonni \thanks{guido.consonni@unicatt.it}}
\affil[1,2]{Department of Statistical Sciences, Universit\`{a} Cattolica del Sacro Cuore, Milan}


\date{}




\maketitle

\begin{abstract}
Our motivation stems from current medical research aiming at personalized treatment using a molecular-based approach. The broad goal is to develop a more precise
and targeted decision making process, relative to traditional treatments based primarily on clinical diagnoses. Specifically, we consider patients affected by Acute Myeloid Leukemia (AML), an hematological cancer characterized by uncontrolled proliferation of hematopoietic stem cells in the bone marrow. Because AML responds poorly to chemoterapeutic treatments, the development of targeted therapies is essential to improve patients' prospects. In particular, the dataset we analyze contains the levels of proteins involved in cell cycle regulation and linked to the progression of the disease. We analyse treatment effects within a causal framework represented by a Directed Acyclic Graph (DAG) model, whose vertices are the protein levels in the network. A major obstacle in implementing the above program is however represented by individual heterogeneity. We address this issue through a Dirichlet Process (DP) mixture of Gaussian DAG-models where both the graphical structure as well as the allied model parameters are regarded as uncertain. Our procedure determines a clustering structure of the units reflecting the underlying heterogeneity, and produces subject-specific estimates of causal effects based on Bayesian model averaging. With reference to the AML dataset, we identify different effects of protein regulation among individuals; moreover, our method clusters patients into groups that exhibit only mild similarities with traditional categories based on morphological features.

\vspace{0.7cm}

\end{abstract}

\section{Introduction}

\subsection{Background and motivation}
\label{sec:intro:background:motivation}


Heterogeneity of individual responses to treatment is a pervasive aspect in a variety of clinical domains. An example  is cancer which is not a single disease as it involves subtypes characterized by distinct sets of molecules; this implies that patients will react differently to the same treatment.
Currently the best treatment is  identified based on clinical diagnoses as opposed to a molecular-based approach
which would allow a more precise and targeted approach to decision making, leading to better treatments and eventually
more favorable prognoses.
At present there exists  a variety of methods to identify cancer subtypes
\citep{Wilkerson:et:al:bioinformatics:2010, Shen:et:al:bioinformatics:2009}; however  they cannot establish whether
subtypes show heterogeneous responses toward
a treatment.
Identifying subtypes with heterogeneous treatment effects is a
causal problem that need be addressed
for a more informed decision strategy \citep{Zhang:bioinformatics:2017}.



In this paper we consider a dataset concerning patients affected by Acute Myeloid Leukemia (AML).
AML represents an aggressive hematological cancer characterized by uncontrolled proliferation of hematopoietic stem cells in the bone marrow.
AML responds very poorly to chemoterapeutic treatments, with a 5-year overall survival rate of about 25\%; the development of new targeted treatments therefore represents a key strategy to improve patients’ prospects.
In addition, AML is widely heterogeneous with numerous genetic aberrations. Consequently, knowledge facilitating individualization of targeted therapies under development in AML is sorely needed. Several subtypes of AML have been identified mainly on the basis of morphologic features.
However, interest has been recently focused on developing improved classification schemes that more accurately
explain the AML heterogeneity and in particular its response to therapies.

The dataset analysed here includes protein levels for 256 newly diagnosed AML patients and is provided as a supplement to \citet{Kornblau:et:al:2009}.
Because protein function regulates the phenotypic characteristics of cancer, a proteomic-based classification can provide relevant information for pathogenesis and prognosis of patients.
In this direction tumor profiling accounting for patients heterogeneity can provide insights on the effect of targeted therapies on individualized basis.
We apply our methodology to the AML dataset to investigate the existence of clustering characterized by differential graph structures as well as model parameter configurations. In addition, we evaluate a battery of causal effects on selected proteins in the network whose regulation
has been established to play a crucial role in AML progression and response to therapy.
We then pair clusters with heterogeneity of causal effects, and identify broad  patterns of potential use for targeted therapy.

\subsection{Overview of our model}

In this subsection we provide the broad picture underlying our model, while leaving technical specifications to later sections.
In our setup we have a collection of variables $\{X_1, \ldots, X_q \}$, exemplified by protein expressions in the AML dataset. When we are interested in isolating one of the variables as a response we label it as $Y$. We consider a sample of  size $n$ from  the $q$-dimensional population.
To account for potential heterogeneity of the subjects  we assume that the statistical model is a mixture of
Directed Acyclic Graph (DAG) models. This means that, if we knew the mixture component  of each sample case, its distributional family would factorize according to the corresponding DAG, equivalently  it would be \textit{Markov} with respect to that DAG.
Graphical models \citep{Laur:1996} are very effective for  encoding conditional independencies. In our setting however we view a DAG as a \textit{causal} model \citep{Pear:2000}; in particular this means that the joint \textit{interventional} distribution of the variables,  following a hypothetical intervention on a variable in the system,  would still be Markov relative to the DAG, save for the component subjected to intervention, which will be specified separately.
This  stability assumption  is crucial to define the notion of total causal effect  on $Y$ following an intervention on $X_j$ \citep{Maathuis:Nandy:Review}.


From a modeling perspective, we start by assuming that the joint observational distribution belongs to a  Dirichlet Process (DP) mixture of Gaussian DAG models; see
\citet{Mueller:Mitra:BA:2013} for an overview of DP mixtures.
The mixing measure is over the space of priors on $(\bmu,\bOmega, \D)$ where $\D$ is a DAG and $(\bmu,\bOmega)$ the parameters of a $q$-variate Normal distribution having precision $\bOmega \in {\cal P}_{\D}$, where ${\cal P}_{\D}$ is the space of symmetric and positive-definite matrices constrained by being Markov relative to $\D$.
We take the baseline measure of the DP as $p(\bmu,\bOmega \g \D )p(\D)$, and specify $p(\bmu,\bOmega \g \D )$
as a Normal-DAG-Wishart prior
after a suitable reparameterization involving 
a Cholesky-type  decomposition of $\bOmega$ \citep{cao:et:al:2019}.
We employ a constructive procedure which produces a prior under \textit{any} DAG model based on the elicitation of a \textit{single} standard Normal-Wishart prior on $(\bmu, \bOmega)$, with $\bOmega$ unconstrained.
In this way,  not only is the elicitation procedure drastically simplified,  but
the marginal likelihood of DAGs belonging to the same Markov  class can be shown to be constant (score equivalence),  besides being available in closed form.
Computations for our model are performed  through an MCMC strategy based on a slice sampler \citep{Walker:2007} and
the Partial Analytic Structure (PAS) algorithm of \citet{Godsill:2012}.

Our contribution can be summarized as follows. We provide a Bayesian modeling framework to evaluate heterogeneous causal effects based  exclusively on \textit{observational}, i.e. non-experimental data. 
\black
Specifically, our model:
i) accounts for individual heterogeneity through an infinite mixture of Directed Acyclic Graphs models; 
ii) allows for structure (DAG) and parameter uncertainty; 
iii) determines a clustering structure of the units;
iv)  produces subject-specific causal effects incorporating uncertainty on various aspects of the model through Bayesian model averaging. 
When applied to the AML data
it highlights
that protein regulation produces heterogeneous effects, and identifies cluster of patients potentially benefiting from selective interventions.

\subsection{Related work}

%
%
%


The literature on the analysis of heterogeneous causal effects has been growing lately, especially in the machine learning community where  tree-based methods, in particular random forests, have found natural applications. Works in this area have been carried  out mostly within the potential outcome framework \citep{Rubin:1974}; see for instance \citet{Athey:Imbens:2016}, \citet{Wager:Athey:Jasa:2018},
\citet{Lee:Bargagli-Stoffi:Dominici:arxiv:2020}
and \citet{Hahn:Murray:Carvalho:BA:2020} for a Bayesian viewpoint.

A more decision theoretic approach is presented in \citet{Shpitser:Sherman:proceedingAI:2018}
who discuss identification of personalized effects in causal pathways.
and  explore  settings  where the goal is to map a unit’s characteristics to a treatment tailored to maximize the expected outcome for that unit. This line of approach is also connected to literature spanning from dynamic treatment regime to mediation analysis.





Moving to a Bayesian perspective,
inference on  causal effects within the causal graph framework has been traditionally carried out assuming a homogeneous population, that is the observations are (conditionally) independent and identically distributed  from a zero-mean Gaussian DAG-model; see for instance \citet{Castelletti:Consonni:Biom:2021} and \citet{Castelletti:Consonni:BA:2021}.

So far heterogeneity has been mostly linked  to differential Bayesian  structural learning as in   multiple graphical modeling, where each model is associated to a specific group which is known in advance; see
\citet{Pete:etal:2015} and \citet{Castelletti:et:al:Stat:medicine:2020}
for directed graphs.
Also, \citet{Ni:Stingo:Baladanda:2017} provide an extensions to multi-dimensional graphs.
A previous important attempt to deal explicitly with heterogeneity using a DP mixture of Gaussian graphical models is
\citet{Rodriguez:et:al:2011}. There are however notable differences with respect to our work. First of all they only consider structural and parameter learning for \textit{undirected} graphs; secondly there is no discussion of causal inference. Bayesian nonparametric techniques have  been used in graphical modeling also for robustness purposes: see \citet{Finegold:Drton:AOAS:2011, Finegold:Drton:BA:2014}, and  \citet{Cremaschi:Argiento:et:al:BA:2019} for an extension to more general measures.

Bayesian nonparametric methods for causal inference have been also employed in a \textit {potential outcome} framework; see for instance \citet{Rubin:Jasa:2005}.
In particular, \citet{Roy:et:al:2016:Biostatistics} consider marginal structural models to evaluate the causal effect of a treatment on a survival outcome and allow for heterogeneity by implementing a dependent DP for the response given a set of confounders.
Moreover, \citet{Oganisian:et:al:2021:Biometrics} implement a DP mixture of zero inflated regression models for pathological data exhibiting excesses of zeros; their method allows for prediction, causal effect estimation and clustering of patients into homogeneous groups sharing the same propensity score distribution.

We close this subsection by pointing to a  more foundational line of research that connects invariance, causality and robustness \citep{Buhlmann:StatScience:2020} where an important role is played by heterogeneity; the latter however relates to different \lq \lq environments\rq \rq{}, seen and unseen, and thus has a broader scope  than the notion of heterogeneity  employed in this paper.

\black
\subsection{Structure of the paper}

The organization of the rest of this paper is as follows.
In Section \ref{sec:background} we provide some background results on causal effect estimation based on DAGs.
We then introduce in Section
\ref{sec:DP:mixture:DAGs}
our DP mixture of Gaussian DAG models with particular emphasis on the specification of the baseline prior  distribution for model parameters. Posterior inference  is  discussed in Section \ref{sec:computational}.
In Section \ref{sec:simulations} we conduct extensive simulation experiments to evaluate the proposed method in terms of structural learning, clustering  and causal effect estimation.
Section \ref{sec:application} is entirely devoted to the analysis of the AML data, highlighting the heterogeneity of causal effects as well as the clustering structure.
Finally, Section \ref{sec:discussion} offers a few points for discussion. Some theoretical results on parameter prior distributions and computational details are reported in the Supplementary material.


\black

\section{Background}
\label{sec:background}

\subsection{Directed acyclic graphs and causal effects}

Let $X_1,\dots,X_q$ be a collection of real-valued continuous random variables with joint p.d.f. $f(x_1,\dots,x_q)$.
Let also $\D=(V,E)$ be a Directed Acyclic Graph (DAG), where $V=\{1,\dots,q\}$ is a set of nodes associated to variables $X_1,\dots,X_q$ and $E\subseteq V\times V$ is a set of edges. If $(u,v)\in E$, then $(v,u)\notin E$ and we say that $u$ is a \textit{parent} of node $v$; by converse we say that $v$ is a \textit{child} of $u$. The set of all parents of $v$ in $\D$ is then denoted by $\pa_{\D}(v)$.
Under DAG $\D$ the joint density  factorizes as
\ben
\label{eq:fact:DAG}
f(x_1,\dots,x_q\g\D)=\prod_{j=1}^{q}f(x_j\g\bx_{\pa_{\D}(j)}).
\een
Factorization \eqref{eq:fact:DAG} is often called the Markov property and determines
conditional independence relations among (set of) nodes which  can be read-off from the DAG using graphical criteria. 
We also assume \emph{faithfulness} of $f(\cdot)$ to $\D$ which prescribes that the conditional independencies implied by \eqref{eq:fact:DAG} are exactly those graphically encoded by $\D$.
We remark that faithfulness holds, up to sets of Lebsegue measure zero, under many common families of distributions such as the Gaussian model (Section \ref{sec:Gaussian:DAGs}).

Consider now a (deterministic) \emph{intervention} on variable $X_s$ which consists in setting $X_s$ to the value $\tilde{x}$ and is denoted as $\textnormal{do}(X_s=\tilde{x})$.
The \emph{post-intervention} density is \citep{Pear:2000}
\black
\ben
\label{eq:intervention:distrib}
f(x_1,\dots,x_{q}\g \textnormal{do}(X_s=\tilde{x}))=
\begin{cases}
	\prod\limits_{j=1,j\ne s}^{q}f(x_j\g \bx_{\pa_{\D}(j)})|_{x_s=\tilde{x}}
	& \text{if $x_s=\tilde{x}$},\\
	\quad \, 0
	& \text{otherwise},
\end{cases}
\een
where, importantly, each term $f(x_j\g \bx_{\pa_{\D}(j)})$ in \eqref{eq:intervention:distrib} is the corresponding (pre-intervention) conditional density  of Equation \eqref{eq:fact:DAG}; this implies that the data generating mechanism is \textit{stable} under  intervention, because the latter only affects the local component distribution $f(x_s\g \bx_{\pa_{\D}(s)})$ which is reduced to a point mass on $\tilde{x}$.

In particular, we are interested in evaluating the causal effect of an intervention $\textnormal{do}(X_s=\tilde{x})$ on a response variable $Y$;  by convention we set $X_1=Y$.
The post-intervention distribution of $Y$ is then obtained by integrating \eqref{eq:intervention:distrib} w.r.t. $x_2,\dots,x_q$ which simplifies to
\ben
\label{eq:intervention:distrib:Y}
f(y\g \textnormal{do}(X_s=\tilde{x})) =
\int f(y\g \tilde{x},\bx_{\pa_{\D}(s)}) f(\bx_{\pa_{\D}(s)}) \, d \bx_{\pa_{\D}(s)};
\een
if $Y \notin \pa_{\D}(s)$; on the other hand if
$Y \in \pa_{\D}(s)$ then $f(y\g \textnormal{do}(X_s=\tilde{x})) =f(y)$;
see \citet[Theorem 3.2.2]{Pear:2000}.
Equation \eqref{eq:intervention:distrib:Y} uses the most common \textit{adjustment} set, namely the set of parents of $X_s$; other valid adjustment sets are however possible; see \citet{Witte:Henckel:Maathuis:Didelez:JMLR:2020}.
Also, it is common to summarize the causal effect on $Y$ of an intervention on $X_s$ through the derivative of the  expected value of \eqref{eq:intervention:distrib:Y}
\ben
\label{eq:causal:effect}
\gamma_s := \frac{\partial}{\partial{x}} \vat(Y\g\text{do}(X_s = \tilde{x}))|_{x=\tilde{x}}.
\een
Clearly if  $f(\cdot)$ belongs to some parametric family indexed by
$\btheta \in \Theta_{\D}$
(a parameter specific to the underlying DAG),
the causal effect $\gamma_s$ will be a function of $\btheta$;  accordingly, inference on $\btheta$ will drive inference on $\gamma_s$.

\subsection{Gaussian DAG models}
\label{sec:Gaussian:DAGs}

In the following we focus on \emph{Gaussian} DAG models and assume
\ben
\label{eq:gaussian:DAG:model}
X_1\dots,X_q\g\bmu,\bOmega\sim\N_q(\bmu,\bOmega^{-1}),
\een
where $\bmu=(\mu_1,\dots,\mu_q)^\top \in \Re^{q}$ and $\bOmega\in\mathcal{P}_{\D}$, the set of all symmetric positive definite (s.p.d.) precision matrices Markov w.r.t. $\D$.

\black

Equation \eqref{eq:gaussian:DAG:model} can be alternatively written as a Structural Equation Model (SEM).
\black
Given $\bSigma=\bOmega^{-1}$,  consider the reparameterization
\ben
\label{eq:chol:reparameterization}
\bL_{\prec j \, ]} = \bSigma^{-1}_{\prec j \succ}\bSigma_{\prec j \, ]} ,
\quad
\bD_{jj} = \bSigma_{jj\g \pa_{\D}(j)},
\quad
\eta_j = \mu_j + \bL^{\top}_{\prec j \, ]} \bmu_{\pa_{\D}(j)},
\een
for $j = 1,\dots, q$, where $\bSigma_{jj|\pa_{\D}(j)} = \bSigma_{jj}-\bSigma_{[\, j\succ} \bSigma^{-1}_{\prec j\succ} \bSigma_{\prec j\,]}$,
$\prec j\,] = \pa_{\D}(j)\times j$, $[\, j\succ\, = j \times \pa_{\D}(j)$, $\prec j\succ\, = \pa_{\D}(j)\times \pa_{\D}(j)$.
Parameters $\bL_{\prec j \, ]}$'s correspond to the non-zero elements of a $(q,q)$ 
matrix $\bL$ with all diagonal entries equal to one. Moreover, if we let $\bD$ be a $(q,q)$ diagonal matrix with $(j,j)$-element $\bD_{jj}$ and $\boldsymbol{\eta}=(\eta_1,\dots,\eta_q)^\top$,
the SEM representation of \eqref{eq:gaussian:DAG:model} is given by
\be
\boldsymbol{\eta} + \bL^\top(X_1,\dots,X_q)^\top = \boldsymbol{\varepsilon},
\ee
where $\boldsymbol{\varepsilon} \sim \N_q(\bzero, \bD)$. Equivalently, we can write
\ben
\label{eq:gaussian:DAG:model:fact}
f(x_1,\dots,x_q\g\bmu,\bOmega,\D)=\prod_{j=1}^{q}d\N(x_j\g \eta_j -\bL_{\prec j\,]}^\top \bx_{\pa_\D(j)},\bD_{jj}),
\een
where $d\N(x\g\mu,\sigma^2)$ denotes the Normal density of $\N(\mu,\sigma^2)$.
Equation \eqref{eq:gaussian:DAG:model:fact} represents  the density of a Gaussian DAG model after the reparameterization
$(\bmu,\bOmega)\mapsto(\boldsymbol{\eta},\bL,\bD)$; compare also Equation \eqref{eq:fact:DAG}.
Finally we note that  $\bOmega=\bL \bD^{-1} \bL^\top$. 

\black

Consider now the causal effect of an intervention $\text{do}(X_s = \tilde{x})$ on $Y=X_1$ as defined in Equation
\eqref{eq:causal:effect}.
Under the Gaussian model \eqref{eq:gaussian:DAG:model}, the post-intervention distribution of $Y$ can be written as
\ben
\label{eq:post-intervention-distr-Y}
f(y\g\textnormal{do}(X_s=\widetilde{x}),\bmu,\bOmega, \D)=\int f(y\g \widetilde{x},\bx_{\pa_{\D}(s)},\bmu,\bOmega)
f(\bx_{\pa_{\D}(s)} \g \bmu,\bOmega)  \, d \bx_{\pa_{\D}(s)},
\een
where each density under the integral sign is a suitable Normal. Taking the expectation of
the post-intervention distribution on the left-hand-side of \eqref{eq:post-intervention-distr-Y}, and interchanging the order of integration in the right-hand-side, one obtains
\black
\ben
\label{eq:do:expected:linear}
\vat(Y\g\textnormal{do}(X_s=\widetilde{x}),\bmu,\bOmega,\D)
&=&
\gamma_0+\gamma_s \widetilde{x} +  \bgamma_{\pa_{\D}(s)}^\top \bmu_{\pa_{\D}(s)}\,
\een
so that, using \eqref{eq:causal:effect},
the causal effect of $\textnormal{do}(X_s=\widetilde{x}_s)$ on $Y$ is $\gamma_s$, the coefficient associated to $X_s$ in the conditional expectation of $Y$ given $\bx_{\fa_{\D}(s)}$ with $\fa_{\D}(s)=s\cup \pa_{\D}(s)$.
Therefore, the causal effect $\gamma_s$ can be retrieved from the covariance matrix $\bSigma=\bOmega^{-1}$ as
\ben
\label{eq:causal:effect:gaussian}
\gamma_s = \left[\left[\bSigma_{Y,\fa_{\D}(s)}\right]\left(\bSigma_{\fa_{\D}(s),\fa_{\D}(s)}\right)^{-1}\right]_{1}
\een
where subscript $1$ refers to the first element of the vector, having implicitly assumed that variable \lq \lq $s$\rq\rq{} appears first in the set $\fa_{\D}(s)$.

\black

\section{Dirichlet process mixture of Gaussian DAG models}
\label{sec:DP:mixture:DAGs}

We consider a Dirichlet Process (DP) mixture of Gaussian DAG models so that
\ben
\begin{aligned}
	\label{eq:DP:Gaussian:DAG}
	X_1,\dots,X_q\g H \sim& \,\,
	\int f(x_1,\dots,x_q\g\bmu,\bOmega,\D) \, H(d\bmu, d\bOmega, d\D) \\
	H \sim & \,\, \textnormal{DP}(\alpha_0,M),
\end{aligned}
\een
where $f(x_1,\dots,x_q\g\bmu,\bOmega,\D)$ denotes the density of a Gaussian DAG model defined in \eqref{eq:gaussian:DAG:model:fact}, and
$H$ follows a DP with parameters $\alpha_0$ (precision) and $M$ (baseline), written  $H(\cdot) \sim DP(\alpha_0,M)$.
With regard to the baseline measure  we set
\ben
\label{eq:baseline:gaussian}
M(d\bmu,d\bOmega,d\D) = p(\bmu,\bOmega \g \D)p(\D) \, d\bmu \, d\bOmega \, d\D,
\een
where priors $p(\bmu,\bOmega\g\D)$ and $p(\D)$ 
will be shortly  defined in Section \ref{sec:prior}.

\vspace{0.5cm}

Let now  $\bx_i=(x_{i,1},\dots,x_{i,q})^\top$, $i=1,\dots,n$, be $n$ independent draws from
\eqref{eq:DP:Gaussian:DAG}.
Recall that in a DP mixture each sample $\bx_i$, $i=1, \ldots n$, has potentially a distinct parameter $\btheta_i=(\bmu_i,\bOmega_i,\D_i)$.
Let $K \le n$ be the number of unique values among $\btheta_1,\dots,\btheta_n$ and $\xi_1,\dots,\xi_n$ a sequence of indicator variables, with $\xi_i\in\{1,\dots,K\}$, such that $\btheta_i=\btheta_{\xi_i}^*$.
Denote now with $\bX$ the $(n,q)$ data matrix  obtained by row-binding the individual observations $\bx_i^\top$'s.
It is instructive to write the DP mixture models in terms of the random partition induced by the $\{\xi_i  \}$'s,
\ben
\label{eq:likelihood:DP}
\quad f(\bX\g\xi_1,\dots,\xi_n, K)
=
\prod_{k=1}^K
\left\{
\int
\left[
\prod_{i:\xi_i=k}
f(\bx_i\g\bmu_k^*,\bOmega_k^*,\D_k^*)
\right]
M(d\bmu_k^*,d\bOmega_k^*,d\D_k^*)
\right\}.
\een
Representation
\eqref{eq:likelihood:DP} is easily interpretable.
The model groups observations into
homogeneous classes,  with samples within each class  generated from a standard Gaussian DAG  model.
In practice,  for a given partition,  
$\bX$ is split into $K$ sub-matrices $\bX^{(k)}$, $k=1,\dots,K$, each $\bX^{(k)}$ collecting all observations $\bx_i$ such that $\xi_i=k$.

\subsection{Prior on DAG parameters}
\label{sec:prior}

We now detail our choice of prior distributions $p(\bmu,\bOmega\g\D)$ and $p(\D)$.

For a given DAG $\D$, let $(\bmu,\bOmega)$ be the corresponding parameters, where
$\bmu\in \Re^{q}$, $\bOmega\in\mathcal{P}_{\D}$.
We first consider the reparameterization $(\bmu,\bOmega)\mapsto(\boldsymbol{\eta},\bL,\bD)$ introduced in Section \ref{sec:Gaussian:DAGs}.
Our elictation procedure relies on the method of \citet{Geig:Heck:2002}.
A main feature of this approach is that we only need to specify a prior for
the parameters of a \emph{complete} DAG model, $\N_q(\bmu,\bOmega^{-1})$, with $\bOmega\in\mathcal{P}$  \textit{unconstrained};
the prior for any other (incomplete) DAG is then derived automatically, as we detail in the Supplementary material.
Additionally, and importantly, this procedure guarantees \emph{compatibility} of priors in the sense that  Markov equivalent DAGs are scored with the same marginal likelihood.
Specifically, we show that a proper Normal-Wishart prior, $(\bmu,\bOmega)\sim\N\W(a_{\mu},\boldsymbol{m},a_{\Omega}, \bU)$, leads to the compatible prior
\ben
\label{eq:prior:fact}
p(\boldsymbol{\eta}, \bD, \bL\g \D) &=& \prod_{j=1}^{q}p(\eta_j, \bL_{\prec j\,]}, \bD_{jj}) \nonumber \\
&=& \prod_{j=1}^{q} p(\eta_j\g\bL_{\prec j\,]}, \bD_{jj})
p(\bL_{\prec j\,]} \g \bD_{jj})
p(\bD_{jj})
\een
where
\be
\bD_{jj} &\sim& \textnormal{I-Ga}\left(\frac{1}{2}a^{\D}_j,
\frac{1}{2}\bU_{jj|\pa_{\D}(j)}\right), \\
\bL_{\prec j\,]}\g\bD_{jj} &\sim& \N_{|\pa_{\D}(j)|}\left(-\bU_{\prec j \succ}^{-1}\bU_{\prec j\,]},\bD_{jj} \,\bU_{\prec j \succ}^{-1}\right), \\
\eta_j \g \bL_{\prec j\,]}, \bD_{jj} &\sim& \N\left(m_j + \bL^{\top}_{\prec j\,]}\boldsymbol{m}_{\pa_{\D}(j)},\bD_{jj}/a_{\mu}\right)
\ee
and $a_j^{\D}=a_{\Omega}+|\pa_{\D}(j)|-q+1$.

\vspace{0.5cm}

\subsection{Prior on DAG structures}
\label{sec:prior:DAGs}

Let $\mathcal{S}_q$ be the space of all DAGs on $q$ nodes.
For a given DAG $\D=(V,E)\in\mathcal{S}_q$, let $\bS^{\D}$ be the 0-1 \emph{adjacency matrix} of its skeleton, that is the underlying undirected graph obtained after removing the orientation of all  its edges. Accordingly, for each $(u,v)$-element in $\bS^{\D}$, $\bS_{u,v}^{\D}=1$ if and only if $(u,v)\in E$ or $(v,u)\in E$, and zero  otherwise.
Conditionally on a prior probability of inclusion $\pi\in(0,1)$ we assume
$
\bS_{u,v}^{\D} \g \pi \overset{\textnormal{iid}}\sim \text{Ber}(\pi)
$
for each $u>v$.
Therefore,
\ben
p(\bS^{\D}\g \pi)=\pi^{|\bS^{\D}|} (1-\pi)^{\frac{q(q-1)}{2}-|\bS^{\D}|},
\een
where $|\bS^{\D}|$ is the number of edges in $\D$ (equivalently in its skeleton) and $q(q-1)/2$ corresponds to the maximum number of edges in a DAG on $q$ nodes.

We then proceed hierarchically by assigning $\pi\sim \textnormal{Beta}(a,b)$. Integrating out $\pi$, the resulting prior on $\bS^{\D}$ is
\be
p(\bS^{\D}) = \frac
{\Gamma \left(|\bS^{\D}| + a\right)\Gamma \left(\frac{q(q-1)}{2} - |\bS^{\D}| + b\right)}
{\Gamma \left(\frac{q(q-1)}{2} + a + b\right)}
\cdot
\frac
{\Gamma \left(a + b\right)}
{\Gamma \left(a\right)\Gamma \left(b\right)}.
\ee
A similar prior was introduced by \citet{Scott:Berger:2010} for variable selection in linear models, where it was also shown to account for multiplicity correction.
Finally, we set
\ben
\label{eq:prior:DAG}
p(\D)\propto p(\bS^{\D}), \quad \D\in\mathcal{S}_q.
\een
Hyperparameters $a$ and $b$ can be chosen to reflect a prior knowledge of sparsity in the graph, if available; in the next section we fix for instance $a=1$, $b=(2q-2)/3$,
which is consistent with an expected prior probability of edge inclusion smaller than $0.5$; see also \citet{Pete:Buhl:2014}. The default choice $a=b=1$, which corresponds to $\pi \sim \textnormal{Unif}(0,1)$, can be instead adopted in the absence of substantive prior information.


\vspace{0.5cm}

Finally, the prior on the precision parameter is taken to be $\alpha_0 \sim \textnormal{Gamma}(c,d)$.
Hyperparameters $c,d>0$ control the prior number of clusters \citep{Escobar:West:1995}.
A sensitivity analysis on a grid of values for $c$ and $d$ led to the choice (hereafter employed) $c=3, d=1$,  which results in a moderate expected number of groups,  and a $90\%$ approximate prior credible interval $1<\alpha_0<6$; \black see also \citet{Murugiah:Sweeting:2012} for empirical approaches driving the choice of $c$ and $d$.

\section{Posterior inference: clustering, structural learning and causal effects}
\label{sec:computational}

We implement an MCMC algorithm to sample from the posterior distribution of the DP mixture model \eqref{eq:DP:Gaussian:DAG}.
Our proposal relies on a \textit{slice sampler} \citep{Walker:2007} which is based on the number of explicitly represented mixture components and maintains the structure of a blocked Gibbs sampler. Full details are provided in the Supplementary material.

The output of our MCMC scheme is a collection of $S$ draws approximately  sampled from the (augmented) posterior of $(\boldsymbol{\xi}, \btheta^*)$
where 
$\btheta^*$  is the triple $(\bmu^*, \bOmega^*, \D^*)$.
Specifically, for each MCMC iteration $t=1,\dots,S$ our algorithm returns
the $n$-dimensional vector of individual  allocations
$\boldsymbol{\xi}^{(t)}=\big(\xi_1^{(t)},\dots,\xi_n^{(t)}\big)$, with
$\xi_i^{(t)}\in\big\{1,\dots,K^{(t)}\big\}$ where $K^{(t)}$ is the number of distinct clusters, together
with the collection of $K^{(t)}$ distinct cluster-specific parameters
$\big\{\btheta^{(t)}_{1},\dots,\btheta^{(t)}_{K^{(t)}}\big\}$
\black
From the MCMC output we can construct an $(n,n)$ posterior similarity matrix
$\boldsymbol{S}$ whose $(i,i^\prime)$-element represents the  posterior probability that subjects $i$ and $i^\prime$  belong to the same cluster, namely
\ben
\label{eq:post:similarity}
\widehat{p}(\xi_i=\xi_{i^\prime} \g\bX) = \frac{1}{S}\sum_{t=1}^S\mathbbm{1}\left\{\xi_i^{(t)}=\xi_{i^\prime}^{(t)}\right\}.
\een
The latter can be used to obtain an estimate $\widehat{\boldsymbol{c}}$ of the partition induced by the DP,  e.g. by including subject $i$ and $i^\prime$ in the same cluster whenever $\widehat{p}(\xi_i=\xi_{i^\prime}\g\bX)$ exceeds a given threshold, say 0.5, as we do in the simulation results of Section \ref{sec:simulations}.

MCMC samples may also be used to provide subject-specific estimates
of DAGs and parameters that are needed to estimate causal effects for each individual as set out in the motivations described in Section \ref{sec:intro:background:motivation}.
To this end, we start by defining for each subject $i$ and edge $(u,v)$, $u\ne v$, the
posterior probability of edge inclusion
\ben
\label{eq:post:edge:inclusion}
\widehat{p}_i(u\rightarrow v\g \bX)=
\frac{1}{S}\sum_{t=1}^S
\mathbbm{1}\left\{
(u,v)\in\D^{(t)}_{\xi_i^{(t)}}
\right\}.
\een
where, with a slight abuse of notation,  $\mathbbm{1}\left\{(u,v)\in\D\right\}=1$ if $\D$ contains the edge $u\rightarrow v$, and zero otherwise.
If we include only those edges $u\rightarrow v$ for which  $\widehat{p}_i(u\rightarrow v\g \bX) > w$, the resulting graph can be adopted as a DAG estimate $\widehat{\D}_i$ provided that the graph is acyclic.
In the following we fix $w = 0.5$.

Recall now the definition of causal effect $\gamma_s$ as a function of the precision (inverse-covariance) matrix in Equation \eqref{eq:causal:effect}.
A Bayesian model averaging (BMA) estimate of $\gamma_s$ for individual $i$ can be recovered from the MCMC output as
$\widehat{\gamma}_{s,i}$
\ben
\label{eq:BMA:causal:estimate}
\widehat{\gamma}_{s,i}=
\frac{1}{S}\sum_{t=1}^S
\gamma_{s,i}^{(t)},
\een
where $\gamma_{s,i}^{(t)}$ is computed as in \eqref{eq:causal:effect:gaussian} by setting $\bSigma = \big[\bOmega_{\xi_i}^{(t)}\big]^{-1}$.

\section{Simulations}
\label{sec:simulations}

In this section we evaluate the performance of our method through simulation studies.

\subsection{Settings}
\label{sec:simulations:settings}

We consider settings with $q=20$ nodes and $K=2$ clusters. For  cluster $k \in \{ 1,2\}$ the sample size $n_k$ takes values in  
$\{50,100,200,500\}$, and for each instance we set $n_1=n_2$.
In addition, cluster-specific parameters $\btheta_k=(\bD_k,\bL_k,\boldsymbol{\eta}_k,\D_k)$ are generated under two scenarios.
In \textit{Equal DAGs} (scenario) we randomly generate a sparse DAG $\D_1$ by fixing a probability of edge inclusion equal to $0.1$ and set $\D_2=\D_1$, which implies that the two DAG models are structurally equal; in \textit{Different DAGs} (scenario) we instead generate $\D_1$ and $\D_2$ independently, so that $\D_1$ and $\D_2$ are different in general.
DAG parameters $(\bD_k,\bL_k,\boldsymbol{\eta}_k)$ are generated independently across $k=1,2$ by setting $\bD_k=\bI_q$, while uniformly sampling the non-zero elements of $\bL_k$ in $[-1,-0.1]\cup[0.1,1]$; in addition,
we sample the elements of each $\boldsymbol{\eta}_k$ in the interval $[-b,b]$, with $b\in\{1,2,5\}$.
Intuitively, higher values of $b$ lead to  stronger separation between the means of the two groups, and this should improve cluster identification.
We then set $\bU=\bI_q$, $a_{\mu}=1$, $\boldsymbol{m}=\boldsymbol{0}$, $a_{\Omega}=q$ in the Normal-DAG-Wishart prior  so that the prior is weakly informative because its weight corresponds to a sample of size one;
see also \citet{Rodriguez:et:al:2011} for a comparison.
Furthermore, to favor sparsity, we fix $a=1$, $b=(2q-2)/3$ in the Beta prior on $\pi$ leading to the prior on DAGs \eqref{eq:prior:DAG}.
From further simulation experiments not reported for brevity it also appeared that results are quite insensitive to these hyperparameter choices, expecially for large sample sizes.
Under each scenario we then perform $N=20$ simulations.
Our MCMC scheme is implemented for a number of MCMC iteration $S=25000$, after having assessed its convergence through some pilot runs.
\black

\subsection{Clustering}
\label{sec:simulations:clustering}

\begin{figure}
	\begin{center}
		\begin{tabular}{cc}
			\quad \quad \quad \textit{Equal DAGs} &
			\quad \quad \quad \textit{Different DAGs}
			\vspace{0.4cm}
			\\
			\includegraphics[scale=0.50]{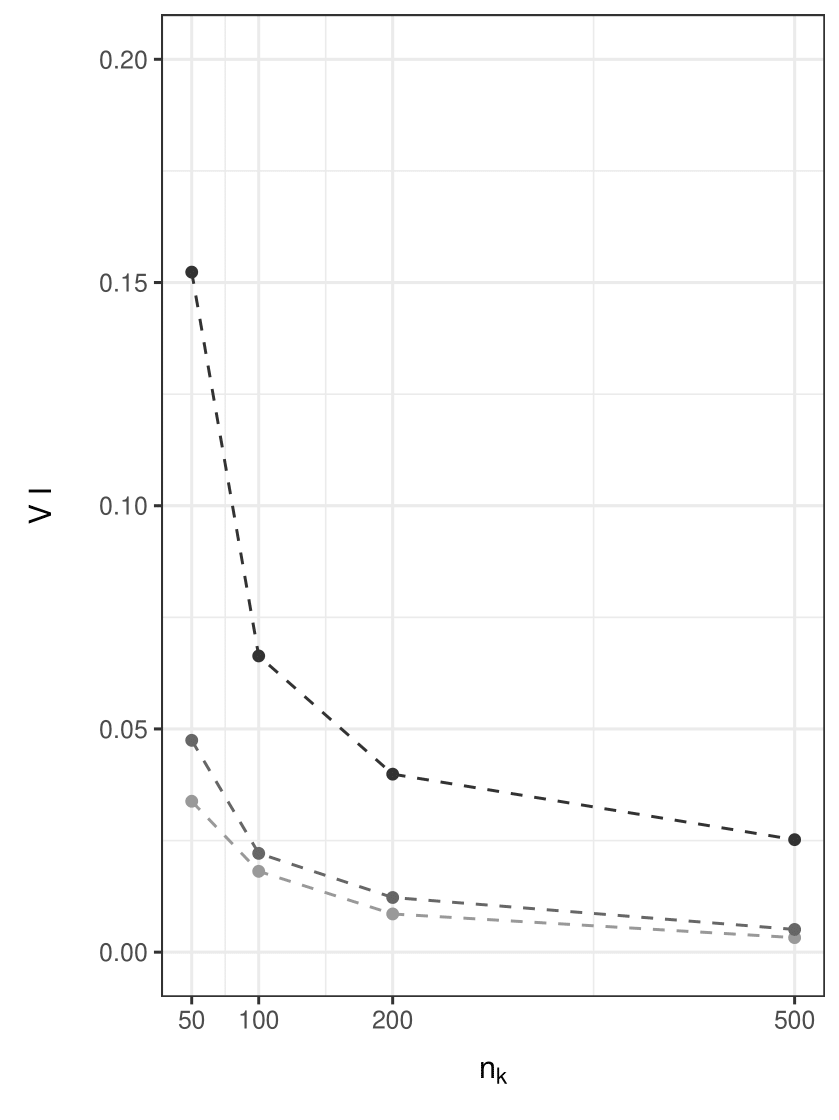}
			&
			\includegraphics[scale=0.50]{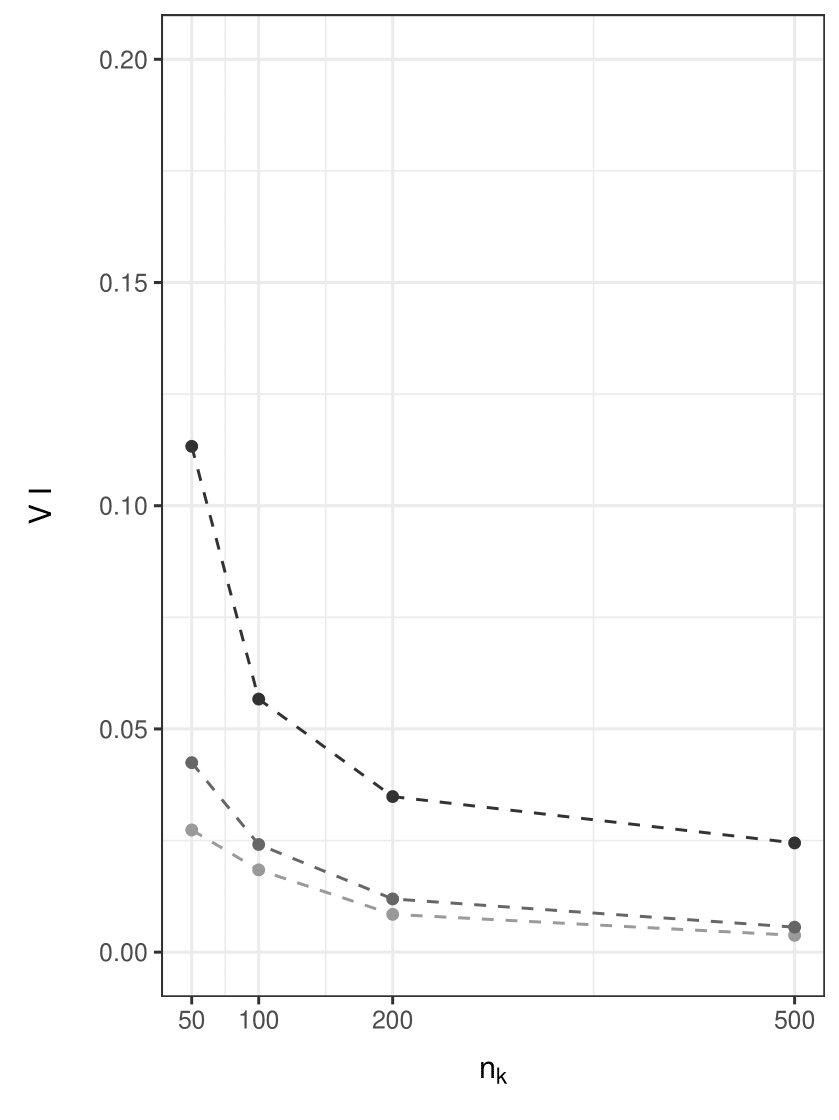}\\
			\includegraphics[scale=0.50]{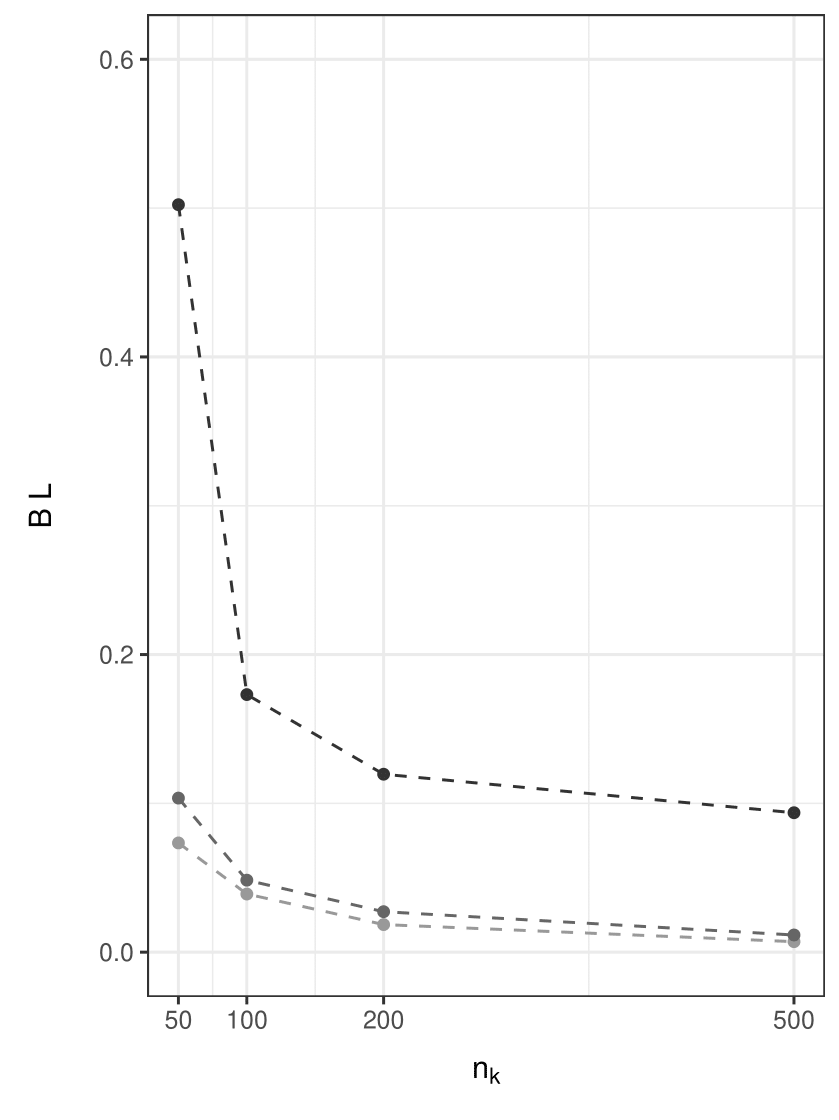}
			&
			\includegraphics[scale=0.50]{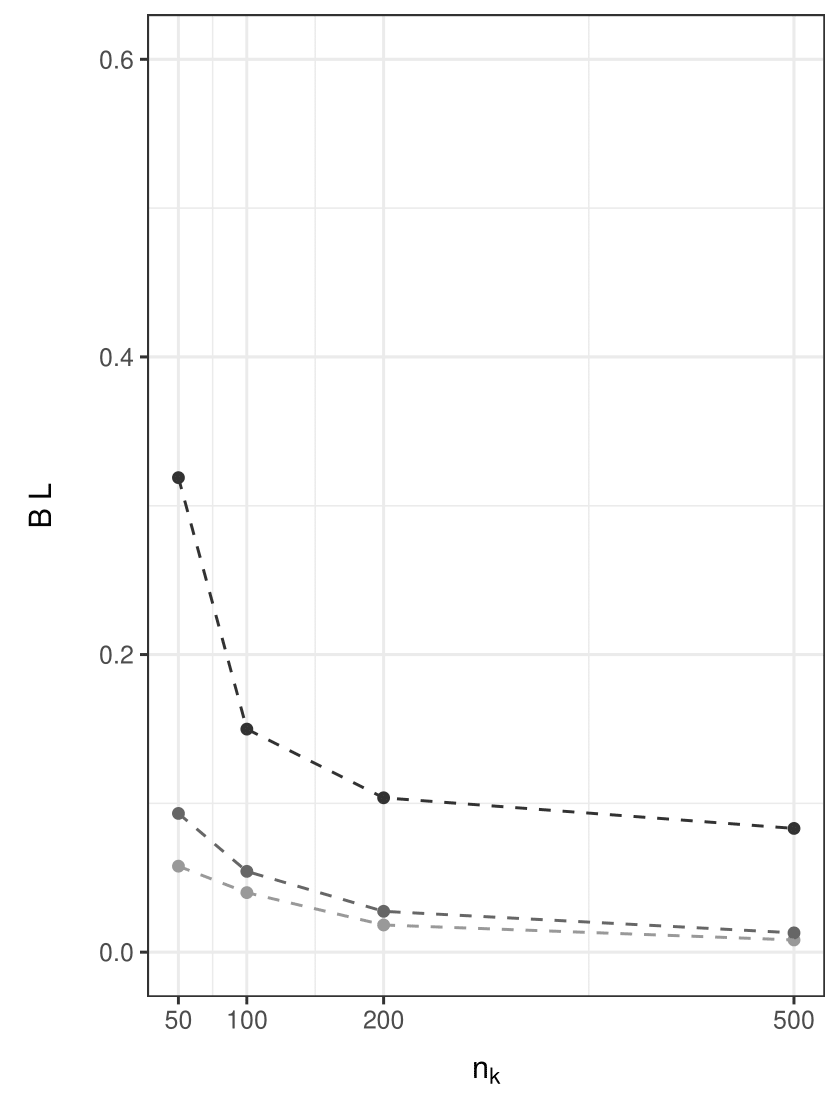}
		\end{tabular}
		\caption{\small Simulations. Average (w.r.t. $20$ simulations) Variation of Information (VI) and Binder Loss (BL) index under \textit{Equal DAGs} and \textit{Different DAGs} scenarios, for increasing sample sizes $n_k\in\{50,100,200,500\}$. Dark, middle and light grey dots correspond to values of $b\in\{1,2,5\}$ respectively.}
		\label{fig:sim:VI:BL}
	\end{center}
\end{figure}

We first evaluate the performance of our method with regard to  cluster allocation.
To this end, we compare the true partition $\boldsymbol{c}$ with the estimated partition $\widehat{\boldsymbol{c}}$ by means of the Binder Loss (BL) \citep{Binder:1978} and the Variation of Information (VI) \citep{Meila:2007}. The two metrics, normalized in $[0,1]$ are respectively defined as
\be
\textnormal{BL}(\boldsymbol{c},\widehat{\boldsymbol{c}}) &=&
\frac{2}{n(n-1)}\sum_{i<j}\big\{
\mathbbm{1}(c_i=c_j)\mathbbm{1}(\widehat c_i\ne\widehat c_j) +
\mathbbm{1}(c_i\ne c_j)\mathbbm{1}(\widehat c_i=\widehat c_j)
\big\}, \\
\textnormal{VI}(\boldsymbol{c},\widehat{\boldsymbol{c}}) &=&
\frac{1}{\log(n)}\big\{H(\boldsymbol{c}) + H(\widehat{\boldsymbol{c}})
- 2I(\boldsymbol{c},\widehat{\boldsymbol{c}})
\big\},
\ee
with $H(\boldsymbol{c})=-\sum_{k=1}^K p(k)\log p(k)$ and
$I(\boldsymbol{c},\widehat{\boldsymbol{c}}) =
\sum_{k=1}^{K}\sum_{h=1}^{H}
p(k,h)\log \{p(k,h)/p(k)p(h)\}$
representing the entropy associated to clustering $\boldsymbol{c}$,
and the mutual information between the two clusterings $\boldsymbol{c}, \widehat{\boldsymbol{c}}$,
where $p(k)=\sum_i \mathbbm{1}(c_i=k)/n$ and $p(k,h) = \sum_i \mathbbm{1}(c_i=k, \widehat{c}_i=h)/n$;
see also \citet{Meila:2007}.
Intuitively, lower values of the two indexes correspond to better performances in the clustering allocation.
We compute BL and VI under each simulated dataset and scenario.
Results are summarized in the plots of Figure \ref{fig:sim:VI:BL}.
Each sequence of points joined by a dotted line represents the average values (w.r.t. the $N=20$ simulations) of an index computed for increasing sample sizes $n_k$ and one value of $b$ (with increasing values of $b$ from dark to light grey).
It appears that the performance of the method improves as $n_k$ grows under each scenario both in terms of BL and VI.
Moreover, higher values of $b$ make  cluster identification easier even for moderate sample sizes, e.g. $n_k=50$.
In addition, the clustering performance is better under \textit{Different DAGs} scenario which corresponds to settings with DAGs generated independently and therefore also ``structurally" different.

\subsection{Structural learning}
\label{sec:simulations:graph}

\black
We now evaluate the performance of our method in learning the graph structures.
To this end, under each simulation, we measure the Structural Hamming Distance (SHD) between each (individual) estimated DAG $\widehat{\D}_i$, $i=1,\dots,n$ and the corresponding true DAG. SHD corresponds to the number of edge insertions, deletions or flips needed to transform the estimated DAG into the true one; accordingly, lower values of SHD correspond to better performances.

For comparison purposes, we also include two alternative,  yet opposed,  learning strategies which are not based on DP mixture models.
The first one corresponds to an oracle setting wherein the true two-group clustering  is known beforehand. We call this benchmark \textit{Two-group oracle}. The second instead wrongly assumes that all observations are conditionally iid from the same one-component model, and we name it \textit{One-group naive}.
Both benchmarks try to evaluate differential performance in structural learning:  the former assesses  the gain afforded by removing  imperfect knowledge on  clustering; the latter   instead captures  decay due to naively neglecting heterogeneity.
In both benchmark strategies,  while not running  a DP mixture model, we use the same specifications for the prior on DAG- and  parameter space.

\begin{figure}
	\begin{center}
		\begin{tabular}{ccc}
			& \quad \,  \textit{Equal DAGs} &
			\vspace{0.4cm}
			\\
			\includegraphics[scale=0.45]{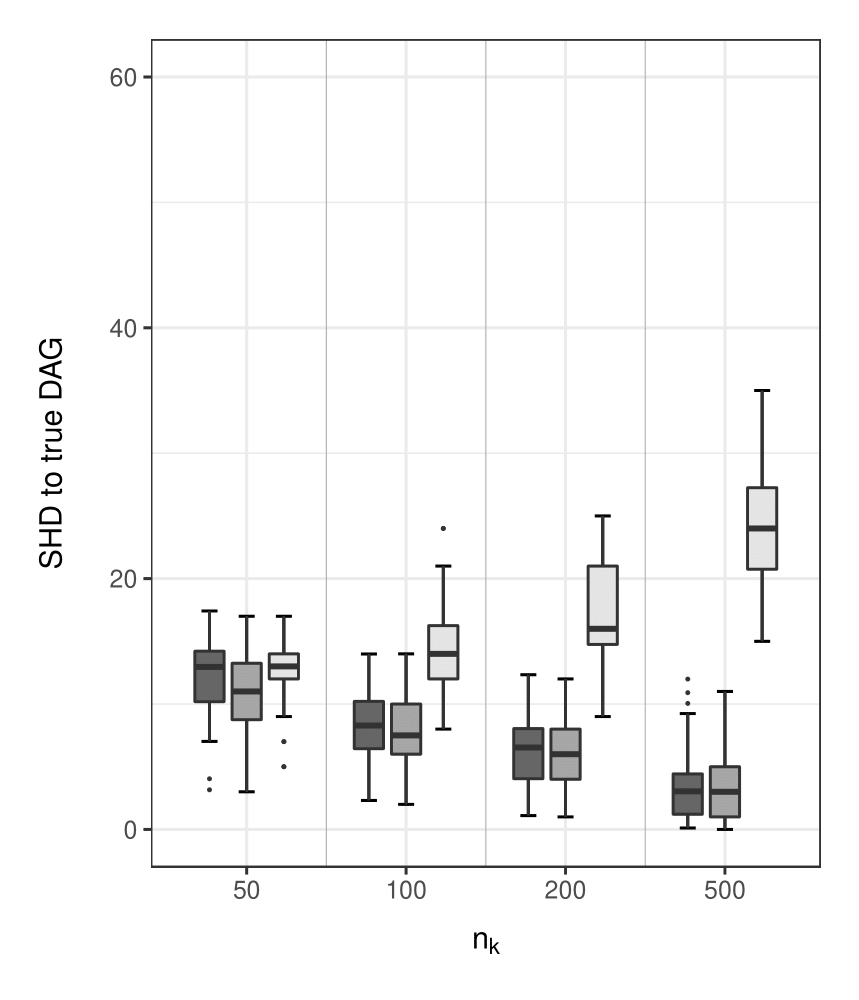} &
			\includegraphics[scale=0.45]{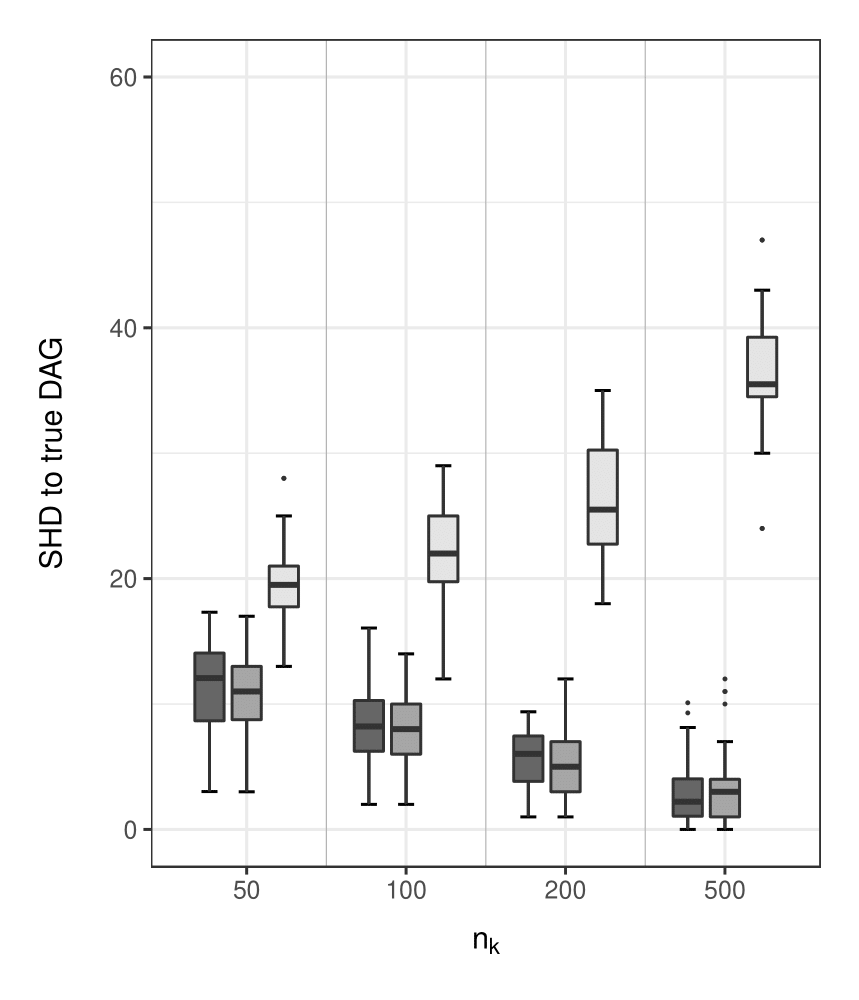} &
			\includegraphics[scale=0.45]{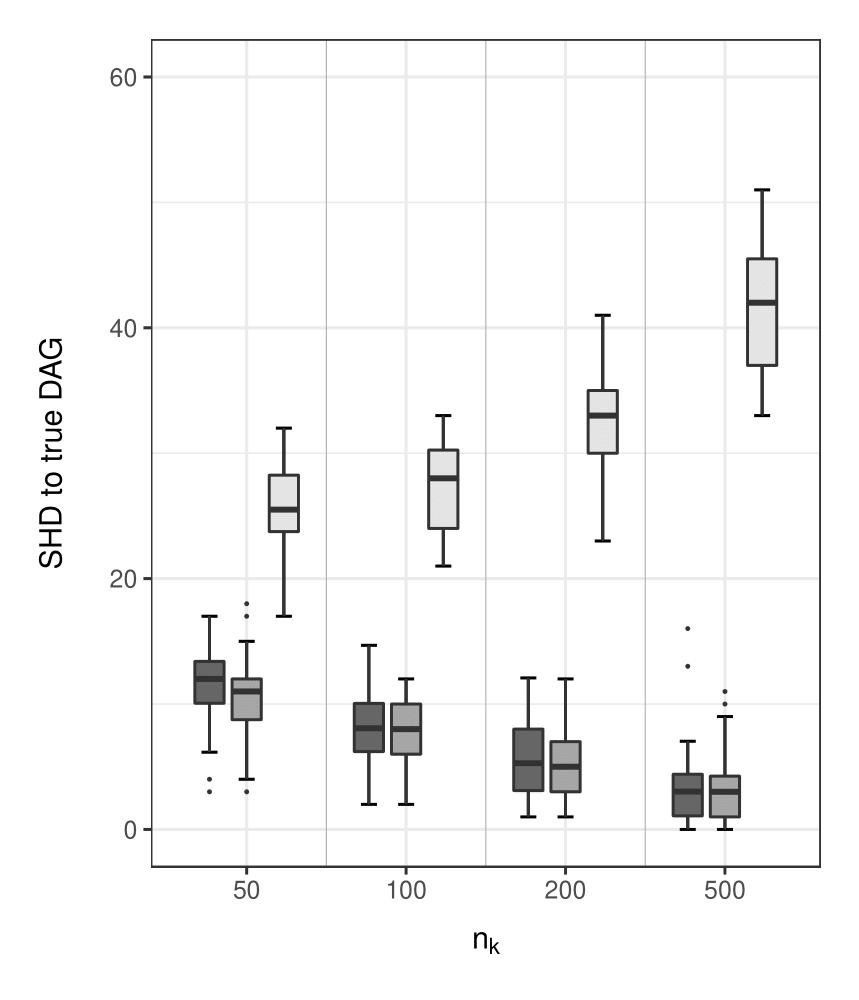} \\
			\vspace{0cm}\\
			& \quad \,  \textit{Different DAGs} &
			\vspace{0.4cm}
			\\
			\includegraphics[scale=0.45]{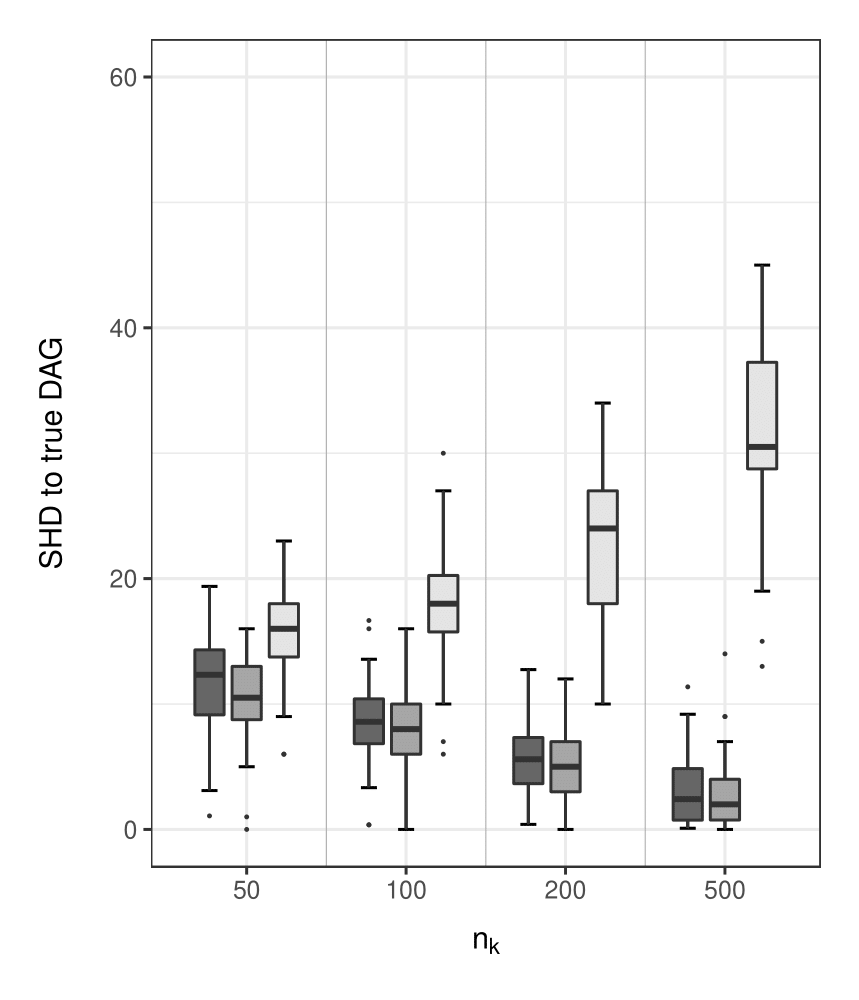} &
			\includegraphics[scale=0.45]{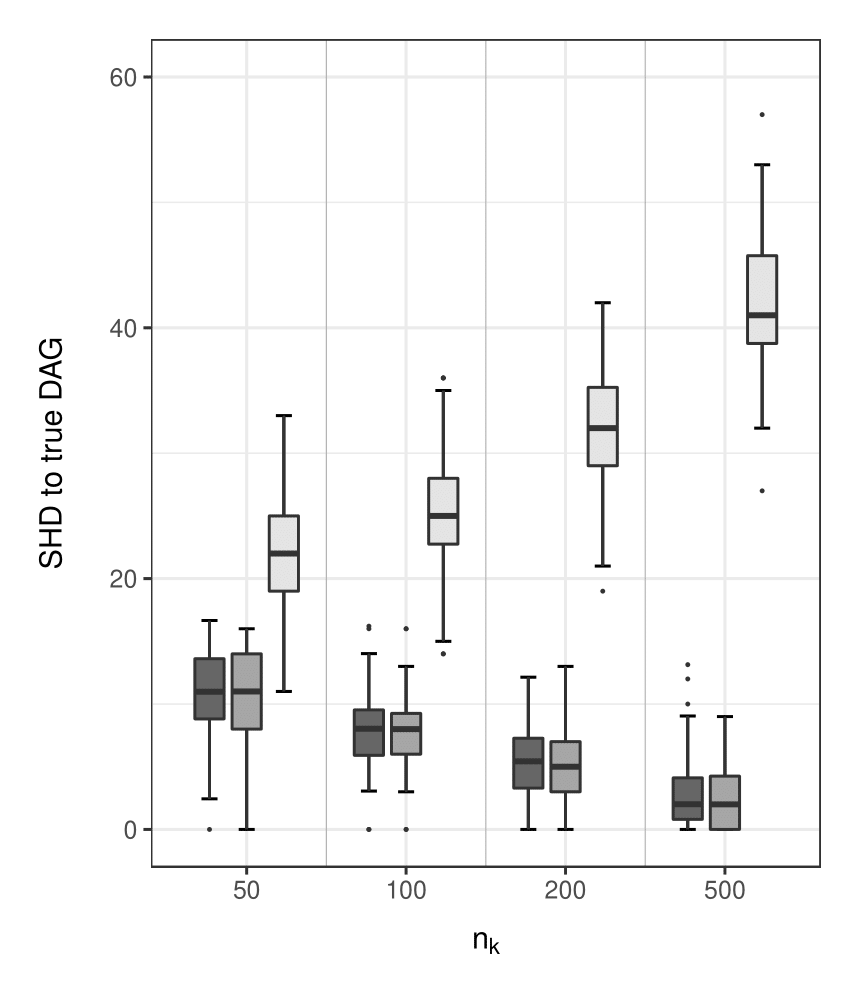} &
			\includegraphics[scale=0.45]{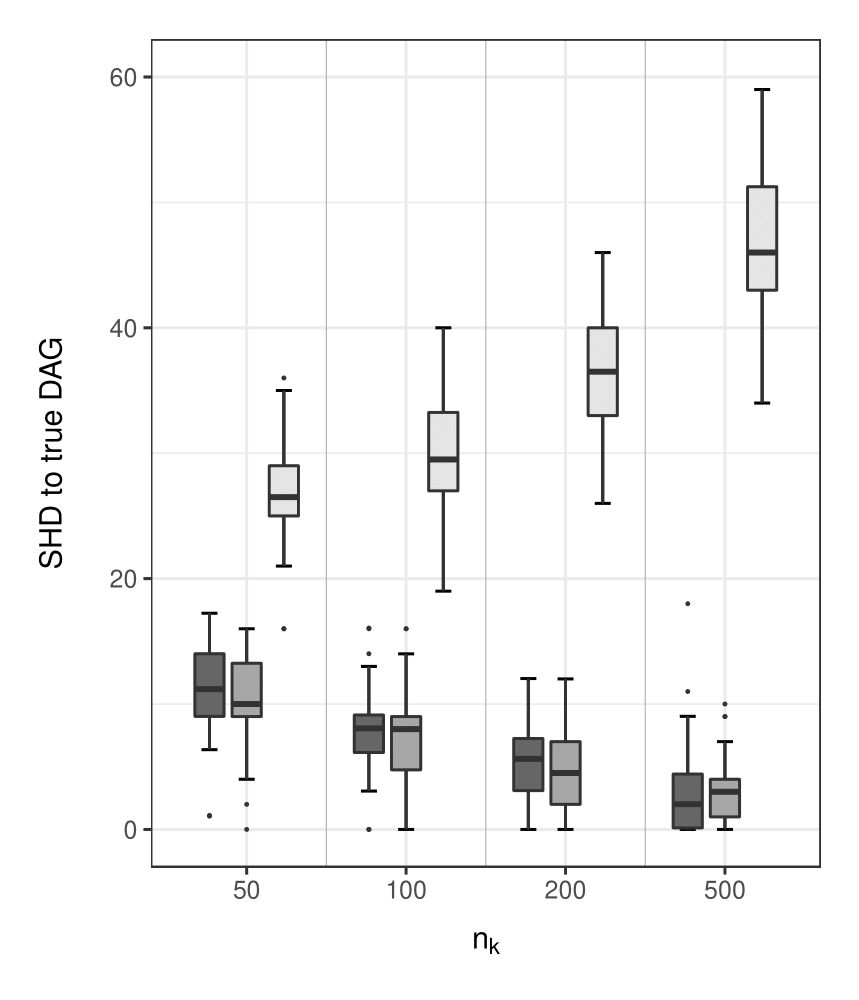} \\
		\end{tabular}
		\caption{\small Simulations. Structural Hamming Distance (SHD) between estimated and true DAGs under \textit{Equal DAGs} and \textit{Different DAGs} scenarios, for increasing sample sizes $n_k\in\{50,100,200,500\}$ and increasing values of $b\in\{1,2,5\}$ (from left to right panels). Dark, medium and light grey box-plots correspond to \textit{DP mixture}, \textit{Two-group oracle} and \textit{One-group naive} strategies respectively.}
		\label{fig:sim:SHD}
	\end{center}
\end{figure}

Results, for each of the two scenarios \textit{Equal DAGs} and \textit{Different DAGs}, are summarized in the plots of Figure \ref{fig:sim:SHD}.
Each box-plot represents the distribution of SHD, averaged with respect to individuals belonging to the same true cluster, 
with increasing
group sample size $n_k\in\{50,100,200,500\}$, and
increasing values of $b\in\{1,2,5\}$ (from top to bottom row). 
It appears that \textit{One-group naive} 
performs worse w.r.t.~the other two methods under both scenarios. In particular its performance worsens as $n_k$ increases and for larger values of $b$, because under both circumstances the two clusters become better separated.
Reassuringly, \textit{Two-group oracle} performs only slightly better than our \textit{DP mixture} method even in the setting $b=1$, where cluster identification is more difficult, with results nearly indistinguishable for $n_k\in\{200,500\}$.
In addition, both methods improve their performance as $n_k$ grows.

\subsection{Causal effect estimation}
\label{sec:simulations:causal}

We finally consider  causal effect estimation.
Under each scenario, we compare the collection of subject-specific BMA causal effect estimates
$\widehat{\gamma}_{s,i}$, $i=1,\dots,n$, $s=1,\dots,q$, in Equation \eqref{eq:BMA:causal:estimate} with the corresponding true causal effects $\gamma_{s,i}$ by means of the absolute-value distance
\be
d_{s,i} = \g\widehat{\gamma}_{s,i} - \gamma_{s,i} \g.
\ee
The distribution of $d_{s,i}$ across subjects and simulated datasets is summarized through the average distance whose percentage values, computed under each scenario, are reported in Tables \ref{tab:distance:equal} and \ref{tab:distance:different}; the two tables refer to scenarios \textit{Equal DAGs} and \textit{Different DAGs} respectively.
For comparisons, we also compute the same collection of causal effect estimates under \textit{Two-group oracle}.
It appears that, while both methods improve their performances as $b$ and $n_k$ grows, \textit{DP mixtures} performs only slightly worse than \textit{Two-group oracle}.
In addition, such differences are more evident under scenarios with moderate sample size $n_k$ and smaller values of $b$, e.g. $b=1$, where indeed cluster allocation was also more difficult.

\begin{table}
	\small
	\vspace{0.1cm}
	\centering
	\begin{tabular}{cccccc}
		\hline
		\hline
		& $n_k$ & 50 & 100 & 200 & 500 \\
		\hline
		\hline
		\multirow{2}{1.5cm}{$b = 5$}
		& \textit{DP mixture} & 3.18 & 2.75 & 2.15 & 1.60 \\
		& \textit{Two-group oracle} & 2.72 & 2.56 & 2.00 & 1.47 \\
		\hline
		\multirow{2}{1.5cm}{$b = 2$}
		& \textit{DP mixture} & 3.41 & 2.90 & 2.25 & 1.72 \\
		& \textit{Two-group oracle} & 2.80 & 2.70 & 2.05 & 1.63 \\
		\hline
		\multirow{2}{1.5cm}{$b = 1$}
		& \textit{DP mixture} & 3.65 & 2.91 & 2.25 & 1.63 \\
		& \textit{Two-group oracle} & 2.90 & 2.56 & 2.04 & 1.41 \\
		\hline
		\hline
	\end{tabular}
	\caption{\label{tab:distance:equal} Simulations. \textit{Equal DAGs} scenario. Average absolute-value distance (computed across simulations and subjects) between estimated and true subject-specific causal effect. Results reported for values of $n_k\in\{50,100,200,500\}$ and $b=\{1,2,5\}$.}
\end{table}

\begin{table}
	\small
	\vspace{0.1cm}
	\centering
	\begin{tabular}{cccccc}
		\hline
		\hline
		& $n_k$ & 50 & 100 & 200 & 500 \\
		\hline
		\hline
		\multirow{2}{1.5cm}{$b = 5$}
		& \textit{DP mixture} & 3.44 & 2.91 & 2.66 & 1.70 \\
		& \textit{Two-group oracle} & 3.10 & 2.63 & 2.60 & 1.67 \\
		\hline
		\multirow{2}{1.5cm}{$b = 2$}
		& \textit{DP mixture} & 3.83 & 2.97 & 2.64 & 1.80 \\
		& \textit{Two-group oracle} & 3.18 & 2.66 & 2.60 & 1.64 \\
		\hline
		\multirow{2}{1.5cm}{$b = 1$}
		& \textit{DP mixture} & 4.19 & 3.06 & 2.74 & 1.90 \\
		& \textit{Two-group oracle} & 3.00 & 2.64 & 2.59 & 1.79 \\
		\hline
		\hline
	\end{tabular}
	\caption{\label{tab:distance:different} Simulations. \textit{Different DAGs} scenario. Average absolute-value distance (computed across simulations and subjects) between estimated and true subject-specific causal effect. Results reported for values of $n_k\in\{50,100,200,500\}$ and $b=\{1,2,5\}$.}
\end{table}

\section{Analysis of AML data}
\label{sec:application}

In this section we apply our methodology to the protein dataset of patients affected by Acute Myeloid Leukemia (AML) described in the Introduction.
This dataset contains the level of $q=18$ proteins and phosphoproteins involved in apoptosis and cell cycle regulation according to the KEGG database \citep{Kanehisa:et:al:2012} for 256 newly diagnosed AML patients and is provided as a supplement to \citet{Kornblau:et:al:2009}.
Classification of AML patients is commonly based on the French-American-British (FAB) system and relies on morphologic features, along with flow cytometric analysis of surface marker expression, cytogenetics, and assessment of recurrent molecular abnormalities. In particular, 11 of the FAB subtypes are present in the dataset, besides one group of patients with unknown subtype; see also Table \ref{tab:subtypes}.

\begin{table}
	\footnotesize
	\vspace{0.1cm}
	\centering
	\begin{tabular}{cccccccccccccc}
		\hline
		\hline
		& subtype & M0 & M1 & M2 & M4 & M4EOS & M5 & M5A & M5b & M6 & M7 & RAEBT & Unknown \\
		\hline
		& size & 17 & 34 & 68 & 59 & 9 & 6 & 13 & 9 & 7 & 5 & 5 & 24 \\
		\hline
		\hline
	\end{tabular}
	\caption{\label{tab:subtypes} AML data. AML subtypes defined according to the French-American-British (FAB) system and corresponding number of subjects (size) in the dataset.}
\end{table}

The same dataset was analysed by
\citet{Pete:etal:2015} and \citet{Castelletti:et:al:Stat:medicine:2020}
from a multiple graphical model-perspective. These Authors included in their analysis four AML subtypes for which a reasonable sample size is available: M0 (17 subjects), M1 (34 subjects), M2 (68 subjects), and M4 (59 subjects).
Specifically,  assuming the four groups were given, both papers
developed a Bayesian analysis which allows  potentially common features in graphical structures - undirected in the first paper and directed in the second one - to be shared among groups.
In  the end, both methods revealed strong similarities between groups in terms of the estimated protein-network structures, but at the same time they were able to identify a few protein interactions specific to a given subtype.
In this paper we take a different approach, and
apply our mixture model to the \textit{full} dataset (including all the $n=256$ subjects) without grouping the patients \emph{a priori}, but rather letting the model cluster the observations as it learns the graphical structuret
of protein-interactions. Eventually, we  also evaluate the effect of interventions on proteins in the network at a subject-specific level.

Among the proteins included in the study, AKT belongs to the phosphoinositide 3-kinase (PI3K)-Akt-mammalian target of rapamycin (mTOR) pathway (PI3K-Akt-mTOR pathway) which is one of the intracellular pathways aberrantly up-regulated in AML \citep{Nepstad:et:al:2020}.
Activation of this pathway (e.g. induced by AKT regulation) has been established to play an important role in leukemogenesis.
In addition, targeting the PI3K-Akt-mTOR pathway with specific inhibitors may produce different effects on AML patients, reflecting biological heterogeneity in the intracellular signaling status \citep{Estruch:et:al:2021}.
Because of the role played in AML progression and response to therapy, we therefore consider the AKT protein and phosphoproteins (AKT, AKT.p308, AKT.p473 in the following) as responses of interest for our causal-effect analysis.

We implement the proposed model by  running our MCMC scheme for $S=120000$ iterations, which includes  a burn-in period of $20000$ runs.
We fix hyper-parameters $c=4$, $d=1$ in the Gamma prior on the precision parameter $\alpha_0$, which is consistent with a moderate expected number of clusters $K$.
In addition, we set $\bU=\bI_q$, $a_{\mu}=1$, $\boldsymbol{m}=\boldsymbol{0}$, $a_{\Omega}=q$ in the Normal-DAG-Wishart prior as in the simulation settings of Section \ref{sec:simulations:settings}.
With regard to the Beta prior on the probability of edge inclusion $\pi$ we instead fix $a=1$, $b=10$, which reflects a moderate degree of sparsity in the network.
To assess the convergence of our algorithm we also run independent MCMC chains with results suggesting a highly satisfactory agreement in terms of clustering, graph structure learning and causal effect estimation; see also the Supplementary material for further details.

\begin{figure}
	\begin{center}
		\begin{tabular}{c}
			\includegraphics[scale=0.52]{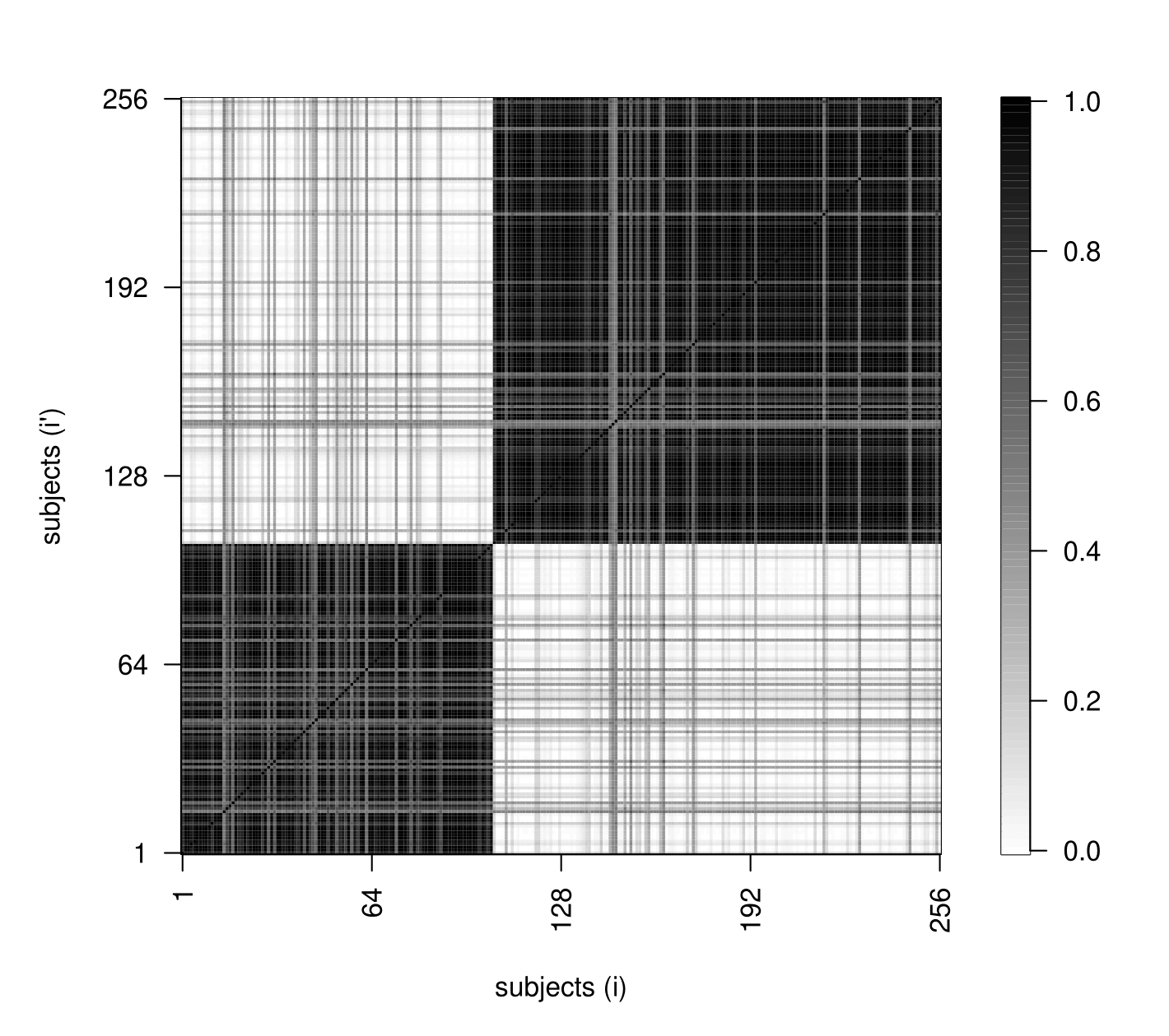}
		\end{tabular}
		\caption{\small AML data. Posterior similarity matrix. Subjects arranged by cluster membership.}
		\label{fig:application:post:similarities}
	\end{center}
\end{figure}

Starting from the MCMC output we first produce the $(n,n)$ similarity matrix based on the posterior probabilities
\eqref{eq:post:similarity} computed for each pair of subjects $(i,i'), i\ne i'$.
In particular,  fixing a threshold for clustering inclusion equal to $0.5$,  we obtain a partition with two groups of size $n_1=105,n_2=151$.
For ease of interpretation we numbered subjects in cluster 1 from 1 to 105, followed by those in cluster 2 from 106 to 256.
Results are summarized in the heat map of Figure \ref{fig:application:post:similarities} where the axes report the ordered subjects.
We note that in both  estimated clusters there exist a few subjects
appearing to be borderline in the sense that they barely qualify for membership to the assigned group
(their inclusion probabilities in groups 1 and 2 are approximatively equal).


\begin{figure}
	\begin{center}
		\begin{tabular}{c}
			\includegraphics[scale=0.55]{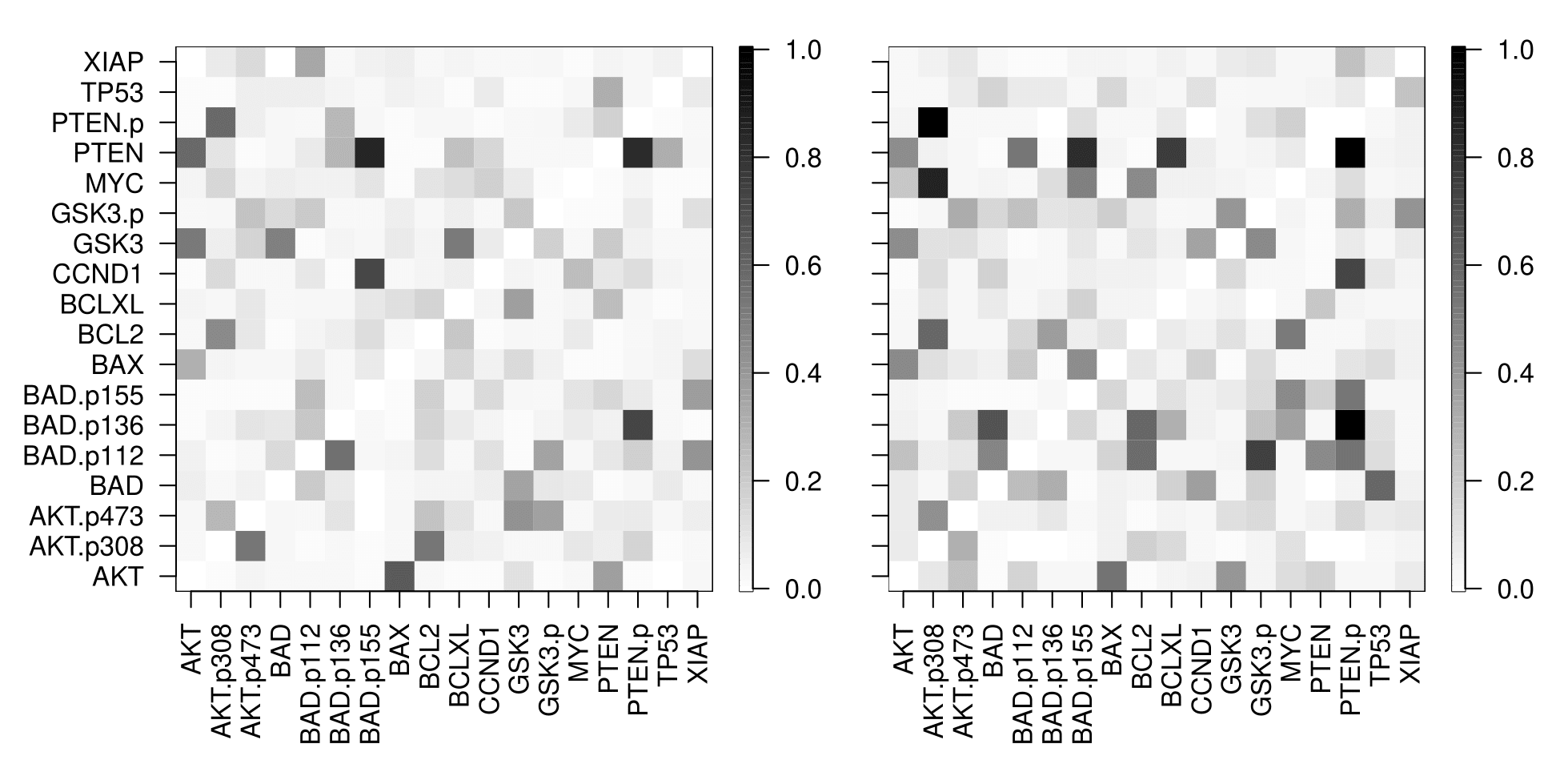}
		\end{tabular}
		\caption{\small AML data. Heat maps of posterior edge inclusion probabilities for two subject-specific graphs. Left map is for one subjects in cluster 1; right map for one subject in cluster 2.}
		\label{fig:application:post:probs:edges}
	\end{center}
\end{figure}

We now focus on graph structure learning. Specifically, we construct for each subject $i=1,\dots,n$ a $(q,q)$ matrix collecting the posterior inclusion probabilities of each (directed) edge  $(u,v)$, $u\ne v$.
Results for two randomly chosen subjects, whose membership is estimated to be cluster 1 and cluster 2 respectively, are reported in Figure \ref{fig:application:post:probs:edges}.
The two heat maps reveal an appreciable degree of sparsity in each of the two
underlying DAG structures, together with some  noticeable differences in the network links.

\begin{figure}
	\begin{center}
		\begin{tabular}{c}
			\includegraphics[scale=0.38]{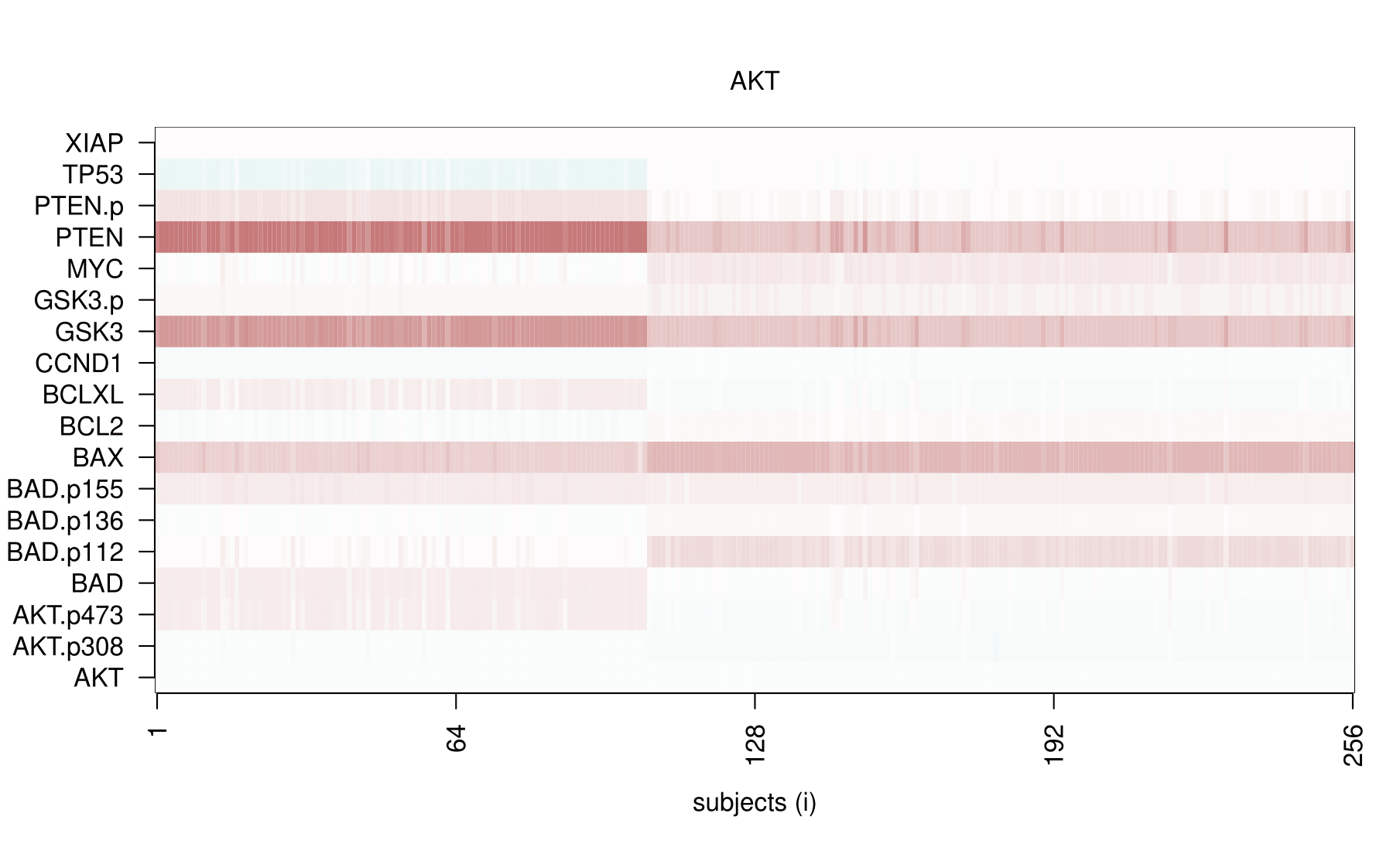}
			\includegraphics[scale=0.38]{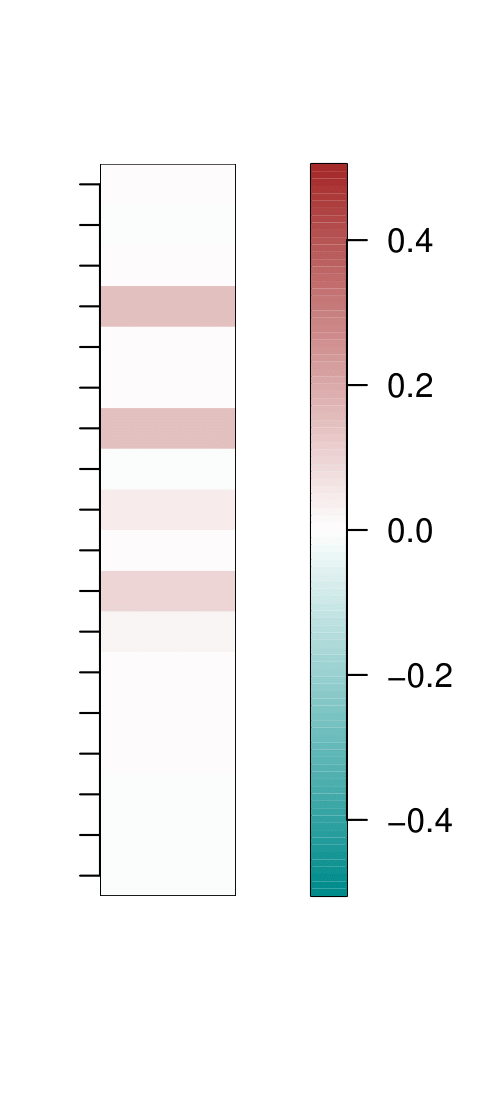} \\
			\includegraphics[scale=0.38]{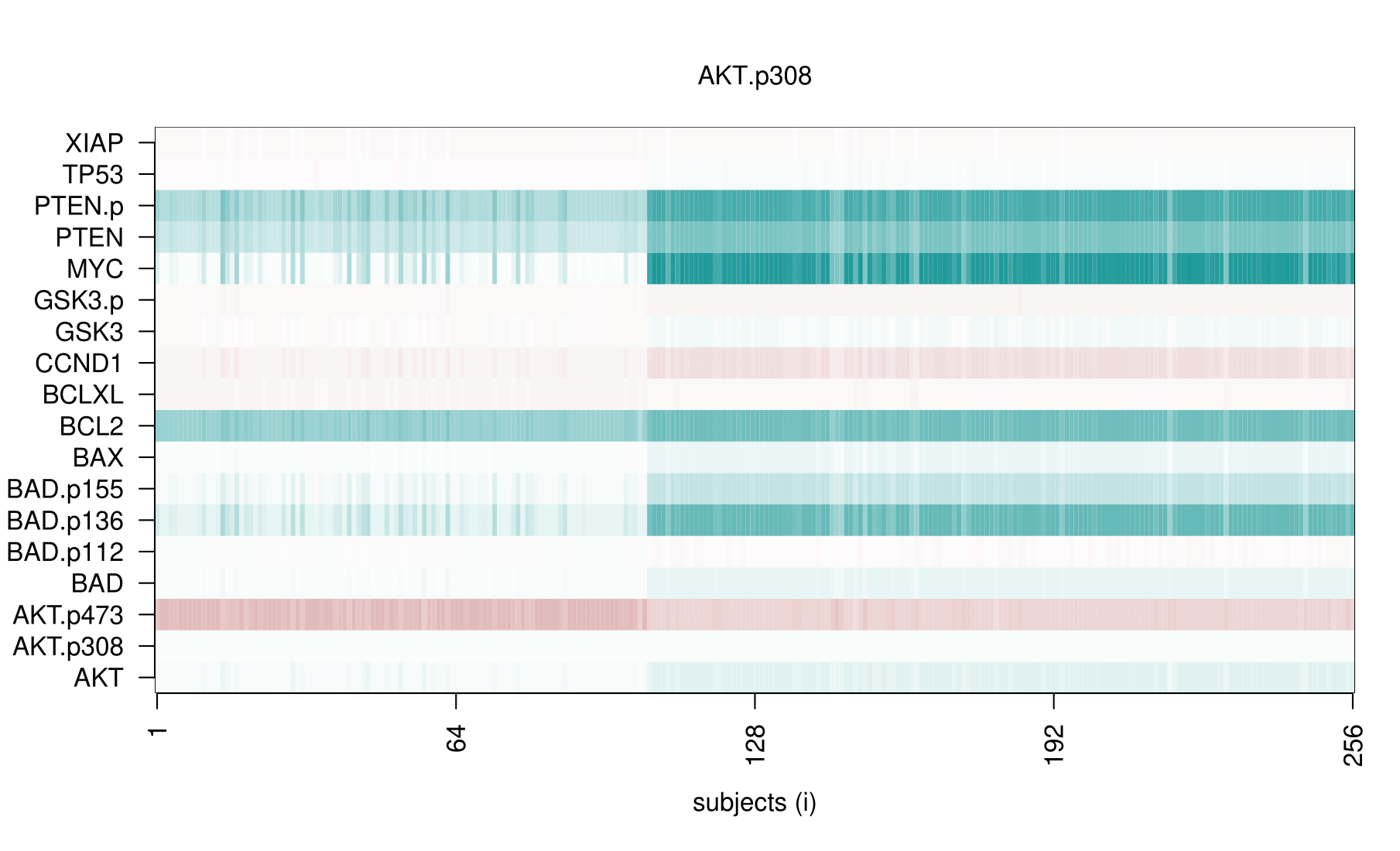}
			\includegraphics[scale=0.38]{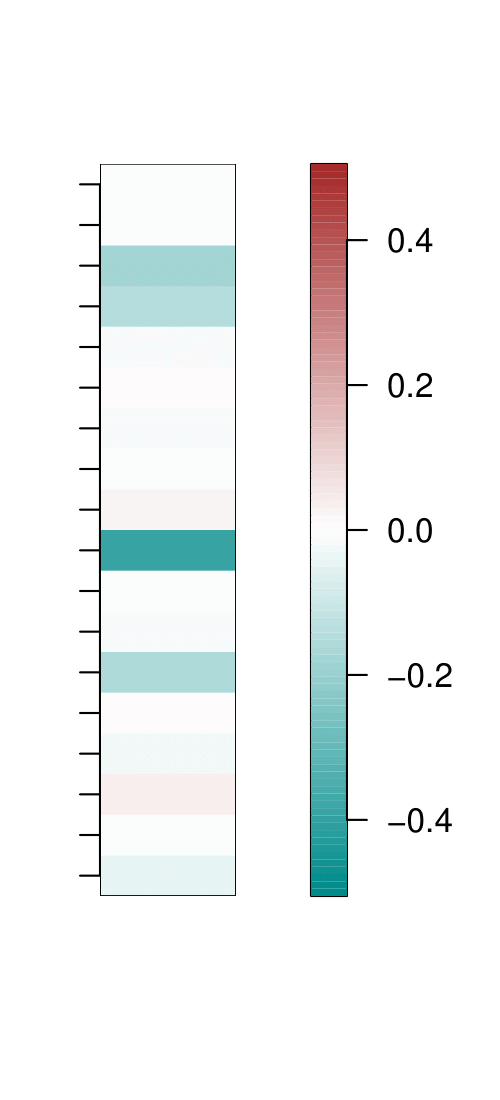} \\
			\includegraphics[scale=0.38]{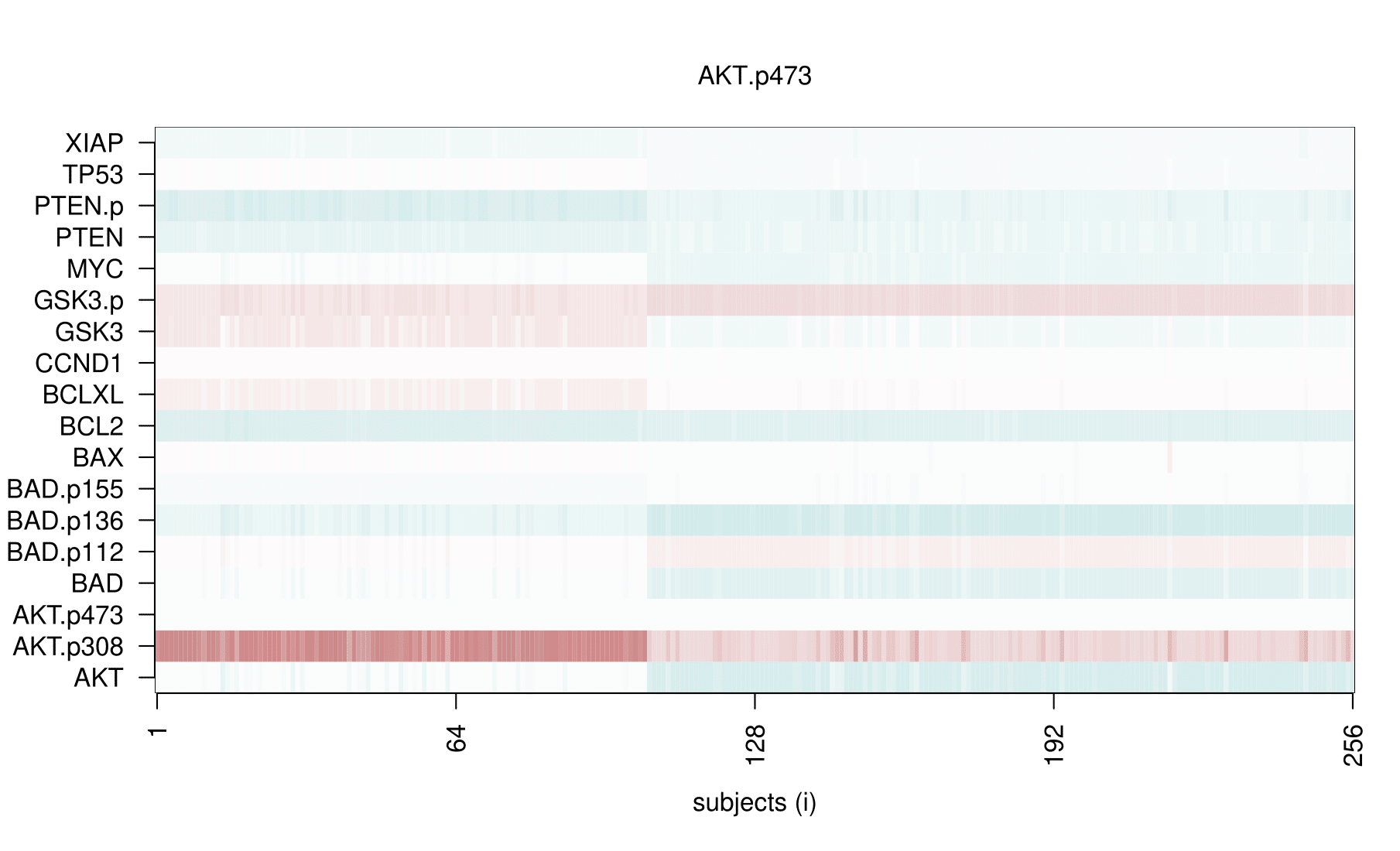}
			\includegraphics[scale=0.38]{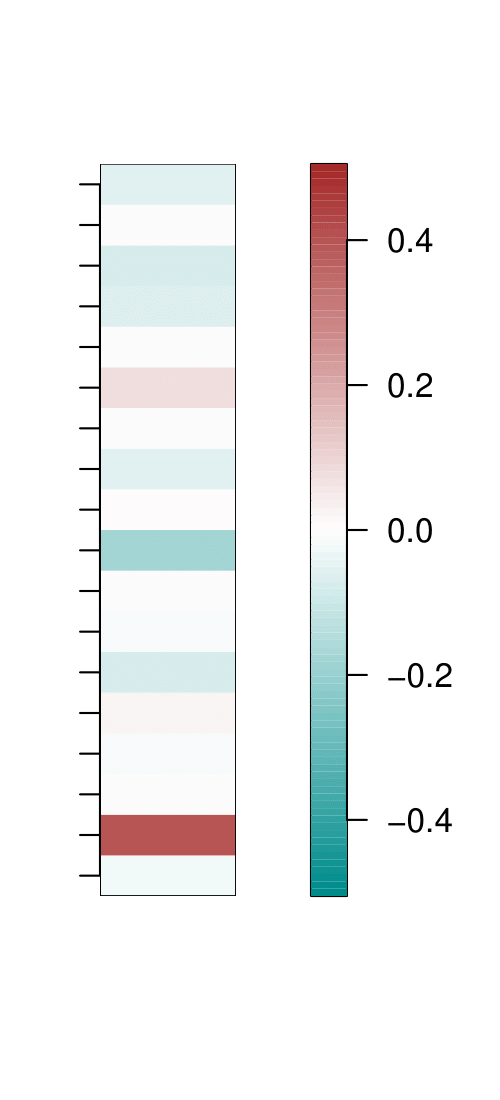}
		\end{tabular}
		\caption{\small AML data. Heat maps of causal effects on responses AKT, AKT.p308, AKT.p473, following an intervention on one target protein among the 18 in the network (AKT, ..., XIAP); left-side heat maps refer to DP-mixture; right-side heat maps to \textit{One-group naive}.}
		\label{fig:application:causal:effects}
	\end{center}
\end{figure}

Using Equation \eqref{eq:BMA:causal:estimate}, we  can provide a subject-specific BMA estimate of the causal effect
on  each of the responses AKT, AKT.p308, AKT.p473 following
an intervention on any other protein $s$ in the network leading to the collection $\gamma_{i,s}$, $s=1,\dots,18$ for each subject $i=1, \ldots, 256$.
Results are summarized in the heat maps of Figure \ref{fig:application:causal:effects} where each plot refers to one of the three response variables.
The pattern already observed in Figure \ref{fig:application:post:similarities} is also apparent. Subjects assigned to the same cluster reveal broadly similar causal effect estimates, with some notable exceptions in both groups for a few individuals.

More interestingly, the effect of an intervention 
varies across the two groups, showing that the cluster structure produced by our analysis has a causal counterpart.
Take for instance the effect of protein PTEN
onto AKT: this is more strongly positive in group 1 than in group 2.
Conversely,  if the response is AKT.p308 the effect in both groups is negative, and more pronounced in group 2.
This finding is of potential interest as protein
PTEN  has been identified as a tumor suppressor because it is capable of  breaking the PI3K-Akt-mTOR pathway,  and therefore represents a common target for inactivation in cancers \citep{Cantley:et:al:1999,Georgescu:et:al:2010}.
From a personalized therapy perspective, one can then  argue that AKT regulation can be induced through  interventions which however should be selected at subject-specific level,
because they can result in heterogeneous causal effects - and corresponding levels of efficacy - across different subjects.
%

\black
In addition, to appreciate the role played by population heterogeneity in causal effect estimation, we compare our results with those based on the alternative \textit{One-group naive} strategy (Section \ref{sec:simulations:graph}).
Results are included in the right-side maps of Figure \ref{fig:application:causal:effects}.

Clearly, in this setting,
causal effects following any intervention  are equal across subjects.
This output reveals substantial differences relative to our previous analysis  suggesting that methods which neglect population heterogeneity can produce misleading estimates of causal effects.
In particular, the one-group assumption has in some cases a ``dilution" effect on coefficients' estimates. This means that each causal effect obtained from \textit{One-group naive} is akin to an  average of  cluster-specific causal estimates which are substantially different among groups. This happens for instance with regard to response AKT for causal effects associated with protein PTEN. Here, the causal effect obtained from \textit{One-group naive} corresponds to a value in between the collection of causal effects resulting from DP mixture. As a consequence the ensuing causal effect coefficient provides an inadequate quantification of the underlying effect because it under- and over- estimates causal effects for individuals in clusters 1 and 2 respectively.


Common  practice  would suggest to first cluster patients into groups according to selected covariates, whenever available, and then infer DAGs and causal effects for each group separately.
In this connection, the leukemia subtypes (Table \ref{tab:subtypes}) may be used for this purpose.
A comparison with this alternative approach is provided in the Supplementary material,
where it is shown that   
patients belonging to  some subtypes, in particular  M0, M1, M2, 
are not primarily assigned to either estimated cluster.  
As a consequence, for subjects belonging to these subtypes there are substantial differences in the resulting causal effect estimates.
This finding suggests that methods which
only rely on a pre-determined group classification may produce inaccurate and possibly misleading estimates of causal
effects.

\black

\section{Discussion}
\label{sec:discussion}

In this paper we  present a Bayesian  framework to evaluate heterogeneous causal effects based on  Directed Acyclic Graphs (DAGs). We model heterogeneity through an infinite mixture model which accounts both for structure  and parameter uncertainty.
The output of our methodology is a collection of subject-specific causal effects, each being a Bayesian model average of causal effects across  DAGs.
Because of the discreteness of the process governing the generation of the individual parameters, a  posterior distribution on the clustering structure of the units is also available.


Our analysis is based on a Dirichlet Process (DP) mixture of Gaussian DAG models. While more general Bayesian nonparametric models might be adopted, e.g. \citet{Mueller:Mitra:BA:2013}, and
\citet{Barrios:etal:2013},
we believe that the main content of our contribution, namely causal inference under heterogeneity based on DAGs, is best captured by the current DP mixture model because of its popularity, interpretability and simplicity of implementation.


Protein and gene expression levels are known to be affected by genetic, environmental, demographic, and other factors. This means that, in addition to the measured variables, there will  typically
be sources of heterogeneity
which are hidden or latent,  and failing to incorporate these sources  may have  detrimental effects on the study
\citep{Leek:Stor:2007}.
Currently  we do not consider latent variables in our model.
In principle they could be made part of our setup,  along the lines of
\citet{Frot:etal:2019} for structure learning, and of
\citet{Shpi:Tche:2016} for the identification of causal effects,
although this would add a significant layer of complexity to the whole procedure.

The type of interventions we have considered may be called \textit{perfect},
meaning that they   eliminate dependencies between targeted variables
and their direct causes; the identifiability of causal DAGs under perfect interventions
was  characterized by
\citet{haus:buhl:2012}.
More recently the  broader notion of \textit{general} intervention has been introduced,
which may modify the dependencies between targeted variables and their causes without
eliminating them; see
\citet{Yang:Katc:uhle:2018}.

\vspace{0.5cm}

{\large \bf Acknowledgments}
This research was partially supported by UCSC (D1 and 2019-D.3.2 research grants). We thank Raffaele Argiento for useful discussions on the slice sampler.

\begin{center}
	{\large\bf SUPPLEMENTARY MATERIAL}
\end{center}
The file \textbf{supplementary\_material.pdf} provides supplemental information to our paper, which is organized into three sections. Section 1 contains a detailed description of the elicitation procedure leading to compatible priors for DAG-model parameters based on Normal-DAG-Wishart distributions.
Section 2 describes our MCMC scheme for posterior inference on DAG structures, cluster allocation parameters and causal effects.
Finally, Section 3 provides convergence diagnostics relative to the application on AML data and additional results.

\bibliographystyle{biometrika} 
\bibliography{biblio}

\end{document}


\title{Supplement to\\Bayesian graphical modeling for heterogeneous causal effects}

\author{Federico Castelletti and Guido Consonni}

\date{}

\maketitle

This supplementary material comprises three sections. Section
\ref{sec:supplementary:prior}
contains a detailed description of the elicitation procedure leading to compatible priors for DAG-model parameters based on Normal-DAG-Wishart distributions.
Section \ref{sec:supplementary:MCMC} describes our MCMC scheme for posterior inference on DAG structures, cluster allocation parameters and causal effects.
Finally, Section \ref{sec:supplementary:application:AML} provides convergence diagnostics relative to our application to AML data together with some additional results.
\black

\section{Prior elicitation for DAG parameters}
\label{sec:supplementary:prior}
For the collection of random variables $(X_1,\dots,X_q)$ consider the Gaussian DAG model
\ben
X_1,\dots,X_q \g \bmu,\bOmega, \D \sim \N_q(\bmu,\bOmega^{-1}),
\een
where $\bmu\in\Re^{q}$ and $\bOmega\in\mathcal{P}_{\D}$, the space of all s.p.d. precision matrices Markov w.r.t. DAG $\D$. To assign a prior on $(\bmu,\bOmega)$ we follow the procedure of \citet{Geig:Heck:2002} (G\&H).
This constructive method assumes some regularity conditions on the likelihood
(\textit{complete model equivalence, regularity, likelihood modularity})
which are satisfied by any Gaussian model.
Starting from representation (7) in our paper for the joint density,
the construction of the prior  is based on two assumptions.
The first one
(\textit{prior modularity}) states that, given two distinct DAG models with the \emph{same} set of parents for vertex $j$, the  prior for the node-parameter $\btheta_j$ must be the same under both models, namely
\be
p(\btheta_j \g \D_h) = p(\btheta_j \g \D_k)
\ee
for any pair of distinct DAGs $\D_h$ and $\D_k$ such that $\pa_{\D_h}(j)=\pa_{\D_k}(j)$.
The second one
(\textit{global parameter independence}) states that for every DAG model $\D$, the parameters $\{\btheta_j; \, j=1, \ldots, q\}$ should be \textit{a priori} independent, that is
\be
p(\btheta\g\D)=\prod_{j=1}^{q}p(\btheta_j\g\D).
\ee
If one follows the above path,  it can be shown that the parameter priors  across all  DAG models are  determined by a \emph{unique} prior on the parameter of \emph{any} of the (equivalent) \emph{complete} DAGs; see Theorem 1 in G\&H.
Additionally all DAG-models within the same equivalence class will be scored equally (same marginal likelihood); see Theorem 3 in G\&H.
G\&H (Section 4) also discuss the Gaussian case.
For the benefit of the reader and for completeness of exposition we provide below the calculations leading to global parameter independence of the Cholesky parameters using our own notation and with some additional details.

%


\subsection{Complete DAG case}

Consider first the case in which $\D$ is complete, so that $\bOmega$ is s.p.d. but otherwise \emph{unconstrained}.
A standard conjugate prior for $(\bmu,\bOmega)$ is  the Normal-Wishart distribution,
$(\bmu,\bOmega)\sim\N\W(a_{\mu},\boldsymbol{m},a_{\Omega}, \bU)$, with $a_{\mu}>0, \boldsymbol{m}\in \Re^q, a_{\Omega}>q-1$ and $\bU$ s.p.d.
Equivalently we can write
\be
\bmu\g\bOmega \sim \N_q\left(\boldsymbol{m}, (a_{\mu}\bOmega)^{-1}\right),
\quad
\bOmega \sim \W_q(a_{\Omega},\bU),
\ee
where our notation for the Wishart means that the density is
$$
p(\bOmega) \propto |\bOmega|^{\frac{a_{\Omega}-q-1}{2}} \exp \left\{ -\frac{1}{2} \tr(\bOmega \bU )  \right\}.
$$
Assume for convenience a \emph{parent ordering} of the nodes which numerically labels the $q$ variables in such a way that $u>v$ whenever $u$ is a parent of $v$.
We introduce the re-parameterization
$(\bmu,\bOmega)\mapsto (\boldsymbol{\eta},\bL,\bD)$, where
$\bL$ is a $(q,q)$ lower triangular matrix with all diagonal entries equal to one and $\bD$ a $(q,q)$ diagonal matrix. In particular, given $\bSigma=\bOmega^{-1}$, we have, for $j=1,\dots, q$,
\ben
\bD_{jj} = \bSigma_{jj\g \pa_{\D}(j)},
\quad
\bL_{\prec j \, ]} = \bSigma^{-1}_{\prec j \succ}\bSigma_{\prec j \, ]} ,
\quad
\eta_j = \mu_j + \bL^{\top}_{\prec j \, ]} \bmu_{\pa_{\D}(j)},
\een
where $\bSigma_{jj|\pa_{\D}(j)} = \bSigma_{jj}-\bSigma_{[\, j\succ} \bSigma^{-1}_{\prec j\succ} \bSigma_{\prec j\,]}$, and
$\prec j\,] = \pa_{\D}(j)\times j$, $[\, j\succ\, = j \times \pa_{\D}(j)$, $\prec j\succ\, = \pa_{\D}(j)\times \pa_{\D}(j)$
Also notice that, because of the parent ordering on $\D$ complete, we have
$\pa_{\D}(j)=\{j+1,\dots,q\}$ and $|\pa_{\D}(j)|=q-j$, for each $j=1,\dots,q$.
We first consider the following Lemma.

%
%

\black
\begin{lemma}
	\label{lemma}
	Let $(\bmu,\bOmega)\sim\N\W_q(a_{\mu},\boldsymbol{m},a_{\Omega}, \bU)$, $a_{\mu}>0, \boldsymbol{m}\in \Re^q, a_{\Omega}>q-1$ and $\bU$ s.p.d.
	Consider the partition $\bx^\top = (\bx_{(1)}^\top \,\bx_{(2)}^\top)$, where $\bx_{(1)}$ is $(q_1,1)$ and $\bx_{(2)}$ is $(q_2,1)$. Partition $\bmu, \bOmega$ and $\bSigma$ accordingly as
	\begin{eqnarray*}
		\bmu = \left(
		\begin{array}{c}
			\bmu_{(1)}\\
			\bmu_{(2)}
		\end{array}\right), \quad
		\bOmega = \left(
		\begin{array}{cc}
			\bOmega_{(1)(1)} & \bOmega_{(1)(2)}\\
			\bOmega_{(2)(1)} & \bOmega_{(2)(2)}
		\end{array}\right), \quad
		\bSigma = \left(
		\begin{array}{cc}
			\bSigma_{(1)(1)} & \bSigma_{(1)(2)}\\
			\bSigma_{(2)(1)} & \bSigma_{(2)(2)}
		\end{array}\right).
	\end{eqnarray*}
	Then,
	$
	\left(\bmu_{(1)},\bOmega_{(1)(1)\g(2)}\right)  \ind \left(\bgamma_{(2)}, \bOmega_{(1)(2)}, \bOmega_{(2)(2)}\right)
	$
	where $\bgamma_{(2)}=\bmu_{(2)}+\bOmega_{(2)(2)}^{-1}\bOmega_{(2)(1)}\bmu_{(1)}$.
	Moreover
	\be
	\left(\bmu_{(1)},\bOmega_{(1)(1)\g(2)}\right) \sim \N\W_{q_1}\left(a_{\mu},\boldsymbol{m}_{(1)}, a_{\Omega}-q_2,\bU_{(1)(1)}\right).
	\ee
\end{lemma}

\begin{proof}
	See G\&H, Theorem 5.
\end{proof}
\black
\vspace{0.5cm}

The following proposition establishes independence among node-parameters $\left(\bD_{jj}, \bL_{\prec j \, ]}, \eta_j\right)$, $j=1,\dots,q$.

\begin{prop}[Global parameter independence]
	Let $(\bmu,\bOmega)\sim\N\W_q(a_{\mu},\boldsymbol{m},a_{\Omega}, \bU)$, $a_{\mu}>0, \boldsymbol{m}\in \Re^q, a_{\Omega}>q-1$ and $\bU$ s.p.d. Consider the re-parameterization
	\be
	\bD_{jj} = \bSigma_{jj\g \pa_{\D}(j)},
	\quad
	\bL_{\prec j \, ]} = -\bSigma^{-1}_{\prec j \succ}\bSigma_{\prec j \, ]} ,
	\quad
	\eta_j = \mu_j + \bL^{\top}_{\prec j \, ]} \bmu_{\pa_{\D}(j)}, \quad
	j=1,\dots,q.
	\ee
	Then  $\ind_{j} \left(\bD_{jj}, \bL_{\prec j \, ]}, \eta_j \right)$;
	in other words the triples $\left(\bD_{jj}, \bL_{\prec j \, ]}, \eta_j \right)$ are mutually stochastically independent.
\end{prop}

\begin{proof}
	Consider first the partition  $\bx^\top= (\bx_{(1)}^\top \bx_{(2)}^\top)$ \black  with $\bx_{(1)} = (x_q,\dots,x_2)^\top$ and $\bx_{(2)}=x_1$
	and for node $j=1$ 
	the re-parameterization
	\be
	\bD_{11} &=& \bSigma_{11\g \pa_{\D}(1)} \\
	\quad
	\bL_{\prec 1 \, ]} &=& -\bSigma^{-1}_{\prec 1 \succ}\bSigma_{\prec 1 \, ]} \\
	\quad
	\eta_1 &=& \mu_1 + \bL^{\top}_{\prec 1 \, ]} \bmu_{\pa_{\D}(1)}.
	\ee
	Equivalently, we can write
	$\bD_{11} = \bOmega_{(2)(2)}^{-1}$ and
	$\bL_{\prec 1 \, ]} = \bOmega_{(2)(2)}^{-1}\bOmega_{(2)(1)}$.\\
	Then, by applying Lemma \ref{lemma} we obtain
	\be
	\left(\bmu_{(1)},\bOmega_{(1)(1)\g(2)}\right)  \ind \left(\bgamma_{(2)}, \bOmega_{(1)(2)}, \bOmega_{(2)(2)}\right).
	\ee
	Moreover, because $\bgamma_{(2)}=\mu_1+\bL_{\prec 1 \, ]}^\top\bmu_{\pa_{\D}(1)}=\eta_1$, we can write
	\be
	\left(\bmu_{(1)},\bOmega_{(1)(1)\g(2)}\right)  \ind \left(\bD_{11}, \bL_{\prec 1 \, ]}, \eta_{1}
	\right).
	\ee

\vspace{0.2cm}

Consider now the partition $\widetilde{\bx}^\top= (\widetilde{\bx}_{(1)}^\top \widetilde{\bx}_{(2)}^\top)$ with $\widetilde{\bx}_{(1)}^\top = (x_q,\dots,x_3)$ and $\widetilde{\bx}_{(2)}=x_2$.
	Similarly as before, we can write
	$\bD_{22} = \widetilde{\bOmega}_{(2)(2)}^{-1}$,
	$\bL_{\prec 2 \, ]} = \widetilde{\bOmega}_{(2)(2)}^{-1}\widetilde{\bOmega}_{(2)(1)}$ and
	$\eta_2 =\mu_2+\bL_{\prec 2 \, ]}^\top\bmu_{\pa_{\D}(2)}$.
	Therefore, from Lemma \ref{lemma} we obtain
	\be
	\left(\widetilde{\bmu}_{(1)},\widetilde{\bOmega}_{(1)(1)\g(2)}\right) \ind \left(\bD_{22}, \bL_{\prec 2 \, ]},
	\eta_{2} \right).
	\ee
Now observe that
$\left(\widetilde{\bmu}_{(1)},\widetilde{\bOmega}_{(1)(1)\g(2)},
\bD_{22}, \bL_{\prec 2 \, ]},\eta_{2} \right)$
is a one-to-one function of 	
$\left(\bmu_{(1)},\bOmega_{(1)(1)\g(2)}\right)$
%
%
whence
	\be
	\left(\widetilde{\bmu}_{(1)},\widetilde{\bOmega}_{(1)(1)\g(2)}\right)
	\ind \left(\bD_{22}, \bL_{\prec 2 \, ]}, \eta_{2} \right)
	\ind \left(\bD_{11}, \bL_{\prec 1 \, ]}, \eta_{1} \right).
	\ee

\vspace{0.2cm}

Proceeding iteratively until
the first block reduces to the null set while the second reduces to $x_q$, so that
$\bD_{qq} = \bSigma_{qq}$,
$\bL_{\prec q \, ]} = \O$,
$\eta_q =\mu_q$,
we obtain
$
\ind_j \left(\bD_{jj}, \bL_{\prec j \, ]}, \eta_{j}\right)
$
which proves global parameter independence.
\end{proof}

\vspace{0.3cm}

\begin{prop}[Node-parameter distribution]
	Let $(\bmu,\bOmega)\sim\N\W_q(a_{\mu},\boldsymbol{m},a_{\Omega}, \bU)$, $a_{\mu}>0, \boldsymbol{m}\in \Re^q, a_{\Omega}>q-1$ and $\bU$ s.p.d. Consider the re-parameterization
	\be
	\bD_{jj} = \bSigma_{jj\g \pa_{\D}(j)},
	\quad
	\bL_{\prec j \, ]} = -\bSigma^{-1}_{\prec j \succ}\bSigma_{\prec j \, ]} ,
	\quad
	\eta_j = \mu_j + \bL^{\top}_{\prec j \, ]} \bmu_{\pa_{\D}(j)}, \quad
	j=1,\dots,q.
	\ee
	Then,
	\be
	\label{eq:prior:parameters:complete}
	\bD_{jj} &\sim& \textnormal{I-Ga}\left(\frac{1}{2}a_j^C,
	\frac{1}{2}\bU_{jj|\pa_{\D}(j)}\right), \\
	\bL_{\prec j\,]}\g\bD_{jj} &\sim& \N_{|\pa_{\D}(j)|}\left(-\bU_{\prec j \succ}^{-1}\bU_{\prec j\,]},\bD_{jj} \,\bU_{\prec j \succ}^{-1}\right), \\
	\eta_j \g \bL_{\prec j\,]}, \bD_{jj} &\sim& \N\left(m_j + \bL^{\top}_{\prec j\,]}\boldsymbol{m}_{\pa_{\D}(j)},\bD_{jj}/a_{\mu}\right)
	\ee
	where $a_j^C=a_{\Omega}+|\pa_{\D}(j)|-q+1$.
\end{prop}

\begin{proof}
	Distributions of $\bL_{\prec j \, ]}$ and $\bD_{jj}$ follow from Theorem 7.1 in
	\citet{Ben:Massam:arXiv}
	where we also used the relationship
	\be
	a_j&=&
	a_{\Omega}+q-2j+3\\
	&=&
	a_{\Omega}+q-2(q-|\pa_{\D}(j)|)+3\\
	&=&
	a_{\Omega}-q+2|\pa_{\D}(j)|+3,
	\ee
	which holds for complete DAGs with a parent ordering of the nodes so that $j=q-|\pa_{\D}(j)|$, and implies
	\be
	\frac{a_j}{2}-\frac{|\pa_{\D}(j)|}{2}-1 = \frac{a_{\Omega}+|\pa_{\D}(j)|-q+1}{2} := \frac{a_j^C}{2}.
	\ee
	Consider now $\eta_j = \mu_j + \bL^{\top}_{\prec j \, ]} \bmu_{\pa_{\D}(j)}$.
	Since $\bmu\g\bOmega \sim \N_q\left(\boldsymbol{m}, (a_{\mu}\bOmega)^{-1}\right)$ and $\eta_j$ is linear in $\mu_j$ and $\bmu_{\pa_{\D}(j)}$, the conditional distribution of $\eta_j$ given $(\bL_{\prec j \, ]}, \bD_{jj})$ is still Gaussian with mean
	\be
	\vat(\eta_j\g \bL_{\prec j \, ]}, \bD_{jj})=
	m_j + \bL^{\top}_{\prec j \, ]} \boldsymbol{m}_{\pa_{\D}(j)}.
	\ee
	In addition,
	\be
	\var(\eta_j\g \bL_{\prec j \, ]}, \bD_{jj})
	=
	\var(\mu_j\g\bD_{jj},\bL_{\prec j \, ]})+
	\var(\bL_{\prec j \, ]}^\top\bmu_{\pa_{\D}(j)})+
	2\textnormal{Cov}(\mu_j,\bL_{\prec j \, ]}^\top \bmu_{\pa_{\D}(j)}).
	\ee
	Using the fact that
	$\textnormal{Cov}(\bA^\top\bx,\bB^\top\by)=\bA^\top\textnormal{Cov}(\bx,\by)\bB$,
	for arbitrary  $(p,1)$ vectors $\bx,\by$ and $(p,p)$ matrices $\bA$, $\bB$ \citep[Equation 8a.1.5]{Rao:1973}
	we can write
	\be
	\var(\eta_j\g \bL_{\prec j \, ]}, \bD_{jj})
	&=&
	\frac{1}{a_{\mu}}
	\left\{
	\bSigma_{jj}+\bL_{\prec j \, ]}^\top\bSigma_{\prec j \succ}\bL_{\prec j \, ]}+
	2\bSigma_{[\, j\succ} \bL_{\prec j \, ]}
	\right\} =
	\frac{1}{a_{\mu}}\bD_{jj},
	\ee
	where we also used the relationships $\bSigma_{jj|\pa_{\D}(j)} = \bSigma_{jj}-\bSigma_{[\, j\succ} \bSigma^{-1}_{\prec j\succ} \bSigma_{\prec j\,]}$ and
	$\bL_{\prec j \, ]} = -\bSigma^{-1}_{\prec j \succ}\bSigma_{\prec j \, ]}$.
	Hence,
	\be
	\eta_j \g \bL_{\prec j\,]}, \bD_{jj} &\sim& \N\left(m_j + \bL^{\top}_{\prec j\,]}\boldsymbol{m}_{\pa_{\D}(j)},\bD_{jj}/a_{\mu}\right).
	\ee
\end{proof}

\subsection{Arbitrary DAGs}

Let now $\D$ be an arbitrary (typically not complete) DAG and assume a parent ordering of its nodes. For each $j \in \{1,\dots,q\}$, let  $\btheta_{j}=\big\{\bD_{jj},\bL_{\prec j \, ]},\eta_j\big\}$ be the parameters associated to node  $j$, and identify a \textit{complete} DAG $\D^{C(j)}$ such that $\pa_{\D^{C(j)}}(j')=\pa_{\D}(j)$, where $j'$ in $\D^{C(j)}$ corresponds to the same variable as $j$ in $\D$. Because of the parent ordering $j'=q-|\pa_\D(j)|$ which is usually different from $j$.
Let
$\btheta_{j'}^{C(j)}$ be the parameter of node $j'$ under the complete DAG $\D^{C(j)}$.
Following the procedure of G\&H we then assign to $\btheta_{j}$ the same prior of $\btheta_{j'}^{C(j)}$ which can be gathered from Equation \eqref{eq:prior:parameters:complete} in the complete Normal-DAG-Wishart version.
In particular, for a given DAG $\D$ we obtain
\ben
\begin{aligned}
	\label{eq:prior:parameters:incomplete:dag}
	\bD_{jj}
	& \,\, \sim \,\,
	\textnormal{I-Ga}\left(\frac{1}{2}a^{\D}_j,
	\frac{1}{2}\bU_{jj\g\pa_{\D}(j)}\right), \\
	\bL_{\prec j\,]}\g\bD_{jj}
	& \,\, \sim \,\,
	\N_{|\pa_{\D}(j)|}\left(-\bU_{\prec j \succ}^{-1}\bU_{\prec j\,]},\bD_{jj} \,\bU_{\prec j \succ}^{-1}\right), \\
	\eta_j \g \bL_{\prec j\,]}, \bD_{jj}
	& \,\, \sim \,\,
	\N\left(m_j + \bL^{\top}_{\prec j\,]}\boldsymbol{m}_{\pa_{\D}(j)},\bD_{jj}/a_{\mu}\right),
\end{aligned}
\een
where
$a_j^{\D}=a_{\Omega}+|\pa_{\D}(j)|-q+1$.
Notice that all distributions in \eqref{eq:prior:parameters:incomplete:dag} only depend on the cardinality of $\pa_\D(j)$ which is the same across alternative parent orderings.
Finally, by assuming independence among node-parameters $(\bD_{jj},\bL_{\prec j \, ]},\eta_j)$, we can write
\ben
\label{eq:prior:dag:fact}
p(\bD,\bL,\boldsymbol{\eta})=\prod_{j=1}^{q}
p(\bD_{jj},\bL_{\prec j \, ]},\eta_j).
\een

\subsection{Posterior distribution}
\label{sec:posterior:distribution}
We now derive the posterior distribution of DAG node-parameters $(\bD_{jj},\bL_{\prec j \, ]},\eta_j)$, $j=1,\dots,q$. For expediency, we proceed by computing first the posterior on the parameters $(\bmu,\bOmega)$ under a complete DAG model $\N_q(\bmu,\bOmega^{-1})$, $\bOmega\in\mathcal{P}$, which by
conjugacy is still Normal-Wishart, and then recover, through the procedure
of G\&H, the posterior on node-parameters.
In particular, given $n$ i.i.d. $q$-dimensional samples $\bx_1,\dots,\bx_n$ collected in the $(n,q)$ data matrix $\bX$, we have
\be
\bmu\g\bOmega,\bX \sim \N_q\left(\boldsymbol{\widetilde{m}}, (\widetilde{a}_{\mu}\bOmega)^{-1}\right),
\quad
\bOmega\g\bX \sim \W_q(\widetilde{a}_{\Omega},\widetilde{\bU}).
\ee
with $\widetilde{a}_{\mu} = a_{\mu}+n$, $\widetilde{a}_{\Omega} = a_{\Omega}+n$, and
\be
\boldsymbol{\widetilde{m}}
&=&
\frac{a_{\mu}}{a_{\mu}+n}\boldsymbol{m}+\frac{n}{a_{\mu}+n}\bar{\bx}, \\
\widetilde{\bU}
&=&
\bU + \bS + \frac{a_{\mu}n}{a_{\mu}+n}\bS_0,
\ee
where $\bS=\sum_{i=1}^n(\bx_i-\bar\bx)(\bx_i-\bar\bx)^\top$,
$\bS_0=(\bar\bx-\boldsymbol{m})(\bar\bx-\boldsymbol{m})^\top$
and $\bar\bx$ is the $(q,1)$ vector collecting the sample means of $X_1,\dots,X_q$.
Therefore, the posterior distribution of the DAG node-parameters $(\bD_{jj},\bL_{\prec j \, ]},\eta_j)$ can be retrieved from \eqref{eq:prior:parameters:incomplete:dag} simply by updating the hyperparameters as $a_{\mu}\mapsto\widetilde{a}_{\mu}$,
$a_{\Omega}\mapsto\widetilde{a}_{\Omega}$, $\boldsymbol{m}\mapsto\widetilde{\boldsymbol{m}}$, $\bU\mapsto\widetilde{\bU}$.

\subsection{Marginal data distribution}

Consider the Gaussian DAG model $X_1,\dots,X_q\g\bmu,\bOmega\sim\N_q(\bmu,\bOmega^{-1})$, $\bOmega\in\mathcal{P}_{\D}$ and the re-parameterization $(\bmu,\bOmega)\mapsto(\bD,\bL,\boldsymbol{\eta})$.
The likelihood function can be written as
\ben
\label{eq:gaussian:DAG:like}
f(\bX\g \bD,\bL,\boldsymbol{\eta},\D)=
\prod_{j=1}^{q}
d\N_n(\bX_j\g \eta_j\mathbf{1}_n -\bX_{\pa_\D(j)}\bL_{\prec j\,]} ,\bD_{jj}\bI_n),
\een
where $\mathbf{1}_n$ is the $(n,1)$ unit vector and $\bI_n$ the $(n,n)$ identity matrix.
Because of parameter prior independence in \eqref{eq:prior:dag:fact} the marginal likelihood of DAG $\D$ admits the same node-by-node factorization, namely
\ben
m(\bX\g\D)
&=&
\int f(\bX\g\bD,\bL,\boldsymbol{\eta},\D) \,p(\bD,\bL,\boldsymbol{\eta}) \, d (\bD,\bL,\boldsymbol{\eta}) \\
&=&
\prod_{j=1}^{q}
m(\bX_j\g\bX_{\pa_{\D}(j)},\D).
\een
In addition, because of conjugacy of the prior $p(\bD_{jj},\bL_{\prec j \, ]},\eta_j)$ with the
Normal density $d\N_n(\bX_j\g\cdot)$, each term $m(\bX_j\g\bX_{\pa_{\D}(j)},\D)$ can be obtained in closed-form expression from the ratio of prior and posterior normalizing constants as
\ben
\label{eq:marg:like:j}
m(\bX_j\g\bX_{\pa_{\D}(j)}, \D) =
(2\pi)^{-\frac{n}{2}}\cdot
\frac{a_{\mu}^{\,\frac{1}{2}}}
{\widetilde{a}_{\mu}^{\,\frac{1}{2}}} \cdot
\frac{\big|\bU_{\prec j \succ}\big|^{\frac{1}{2}}}
{\big|\widetilde{\bU}_{\prec j \succ}\big|^{\frac{1}{2}}}\cdot
\frac{\Gamma\left(\frac{1}{2}\widetilde{a}_j^{\D}\right)}
{\Gamma\left(\frac{1}{2}a_j^{\D}\right)}\cdot
\frac{\Big(\frac{1}{2}\bU_{jj\g\pa_{\D}(j)}\Big)^{\frac{1}{2}a_j^{\D}}}
{\left(\frac{1}{2}\widetilde{\bU}_{jj\g\pa_{\D}(j)}\right)^{\frac{1}{2}\widetilde{a}_j^{\D}}}, \quad \quad
\een
where $\widetilde{a}_j^{\D}=\widetilde{a}_{\Omega}+|\pa_{\D}(j)|-q+1$.

\section{MCMC sampler}
\label{sec:supplementary:MCMC}
In this section we provide full details on the MCMC scheme that we adopt for posterior inference on our model.
For completeness, we also summarize some  basic concepts on DP mixture models.

\begin{definition}
	A random distribution $H$ follows a Dirichlet process (DP) with parameters $\alpha_0$ (precision) and $M$ (baseline), written  $H(\cdot) \sim DP(\alpha_0,M)$, if
	\begin{itemize}
		\item[-] $H(\cdot) = \sum\limits_{k=1}^{\infty}\omega_k \delta_{(\btheta_k^*)}(\cdot)$,
		with $\theta_1^*,\theta_2^*,\dots \overset{\textnormal{iid}}\sim M$,
		\item[-] $\omega_k = v_k\prod\limits_{s < k}(1-v_s)$,
		with $v_1,v_2,\dots \overset{\textnormal{iid}}\sim \textnormal{Beta}(1,\alpha_0)$.
	\end{itemize}
\end{definition}

\noindent
In our setting $\btheta_k^*$ is represented by the triple $(\bmu_k^*, \bOmega_k^*, \D_k^*)$, namely the mean vector and precision matrix $(\bmu_k^*, \bOmega_k^*)$ corresponding to the DAG $\D_k^*$.
Moreover, $\{\omega_k\}_{k=1}^{\infty}$ are weights satisfying $\omega_k\in(0,1)$ and $\sum_{k=1}^\infty\omega_k=1$.
Therefore, sampling from $H$ is equivalent to drawing from the set
$\{\btheta_k^*\}$'s with probability $\omega_k$, $k=1,\dots,\infty$.

Let now $X_1,\dots,X_q$ be a collection of real-valued random variables.
Conditionally on the discrete random measure $H$, we assume that the vector $(X_1,\dots,X_q)$ follows a DP mixture of Gaussian DAG models, namely
\ben
\begin{aligned}
	\label{eq:DP:Gaussian:DAG:suppl}
	X_1,\dots,X_q\g H \sim& \,\,
	\int f(x_1,\dots,x_q\g\bmu,\bOmega,\D) \, H(d\bmu, d\bOmega, d\D) \\
	H \sim & \,\, \textnormal{DP}(\alpha_0,M),
\end{aligned}
\een
where $f(x_1,\dots,x_q\g\bmu,\bOmega,\D)$ denotes the density of a Gaussian DAG model, as defined in Section 2.2 of our paper.
With regard to the baseline measure we set
\ben
M(d\bmu,d\bOmega,d\D) = p(\bmu,\bOmega \g \D)p(\D) \, d\bmu \, d\bOmega \, d\D
\een
with priors $p(\bmu,\bOmega\g\D)$ and $p(\D)$ 
defined in  Section 3.1 of our paper.
Let now  $\bx_i=(x_{i,1},\dots,x_{i,q})^\top$, $i=1,\dots,n$, be $n$ independent draws from
\eqref{eq:DP:Gaussian:DAG:suppl}.
Recall that in a DP mixture each sample $\bx_i$, $i=1, \ldots n$,  has potentially  a distinct parameter $\btheta_i=(\bmu_i,\bOmega_i,\D_i)$.
Let $K \le n$ be the unique values among $\btheta_1,\dots,\btheta_n$ and $\xi_1,\dots,\xi_n$ a sequence of indicator variables, with $\xi_i\in\{1,\dots,K\}$, such that $\btheta_i=\btheta_{\xi_i}^*$.
Denote now with $\bX$ the $(n,q)$ data matrix  obtained by row-binding the individual observations $\bx_i^\top$'s.
The DP mixture model can be written in terms of the random partition induced by the $\{\xi_i  \}$'s
\ben
\label{eq:likelihood:DP}
f(\bX\g\xi_1,\dots,\xi_n, K)
=
\prod_{k=1}^K
\left\{
\int
\left[
\prod_{i:\xi_i=k}
f(\bx_i\g\bmu_k^*,\bOmega_k^*,\D_k^*)
\right]
M(d\bmu_k^*,d\bOmega_k^*,d\D_k^*)
\right\}.
\een

\vspace{0.5cm}

To sample from the posterior of the DP mixture we rely on the \emph{slice sampler} \citep{Walker:2007}; see also \citet{Kalli:et:al:2011}.
Recall now the likelihood function in Equation \eqref{eq:likelihood:DP} and the prior on parameters defined in Sections 3.1 and 3.2,
where in particular $\btheta_k^{*}\overset{iid}{\sim}M$.
We use the auxiliary variables $\left\{v_k\right\}_{k=1}^{\infty}$ such that
$v_k\overset{iid}{\sim}\textnormal{Beta}(1,\alpha_0)$
and
$\omega_k=v_k\prod_{h<k}(1-v_h)$,
where $\omega_k$'s are the weights of the DP; see also Section 3 in the paper.

Let also $u_1,\dots,u_n$ be uniformly distributed auxiliary variables such that
\be
p(u_i,\bx_i,\xi_i\g \bv, \btheta_i^*)
&=&
f(\bx_i\g\btheta_{\xi_i}^*)\,\mathbbm{1}(u_i<\omega_{\xi_i})\\
&=&
f(\xi_i\g\bv)f(u_i\g\xi_i,\bv)f(\bx_i\g\btheta_{\xi_i}^*).
\ee
The \emph{augmented} joint distribution of the data and parameters can be written as
\be
f(\bX,\bu,\bv,\boldsymbol{\xi},\btheta^*, \alpha_0)=
\prod_{i=1}^n
\left\{
f(\xi_i\g\bv)f(u_i\g\xi_i,\bv)f(\bx_i\g\btheta_{\xi_i}^*)
\right\}
\prod_{k=1}^K p(v_k)
\prod_{k=1}^K p(\btheta_k^*)\cdot
p(\alpha_0)
\ee
where $\bu=(u_1,\dots,u_n)^\top$ and $\boldsymbol{\xi}=(\xi_1,\dots,\xi_n)^\top$.
Update of parameters $(\bu,\bv,\boldsymbol{\xi},\btheta^*)$ is performed in the following steps.

\subsection{Update of ($\boldsymbol{u}$,$\boldsymbol{v}$)}

Block-update of parameters $(\bu,\bv)$ is performed by sampling from their joint full conditional
distribution
$
f(\bu,\bv \g \cdot) = f(\bu\g\bv,\cdot)f(\bv\g\cdot).
$
In particular, we have
\be
f(\bu\g\bv,\cdot) \propto \prod_{i=1}^n f(u_i\g\xi_i,\bv)
=\prod_{i=1}^n \mathbbm{1}(u_i<\omega_{\xi_i}),
\ee
so that $u_i$'s are mutually independent a posteriori with distribution
\be
u_i\g\cdot \sim\textnormal{Unif}\big(0,\omega_{\xi_i}\big), \quad i=1,\dots,n.
\ee
Moreover,
\be
f(v_k\g\cdot)
\propto
\prod_{i=1}^n
\big\{
f(\xi_i\g\bv)f(u_i\g\xi_i,\bv)
\big\}
\, p(v_k).
\ee
Recalling that $\omega_k = v_k\prod_{h<k}(1-v_h)$, $v_k\sim \textnormal{Beta}(1,\alpha_0)$ and
$\Pr(\xi_i=k\g\bv)=\omega_{k}$
we can write
\be
f(v_k\g\cdot)
&\propto&
\prod_{i=1}^n
\left\{
v_{\xi_i}
\prod_{h<\xi_i}(1-v_h)
\right\}
(1-v_k)^{\alpha_0-1}\\
&=&
\prod_{k=1}^K
\prod_{i:\xi_i=k}
\left\{
v_k \prod_{v<h}(1-v_h)
\right\}
(1-v_k)^{\alpha_0-1},
\ee
where $K=\max\{\xi_i,i=1,\dots,n\}$.
It is then straightforward to show that $v_1,\dots,v_K$ are mutually independent with distribution
\be
v_k\g \cdots \sim \textnormal{Beta}\left(n_k+1,\alpha_0+\sum_{h>k}n_k\right), \quad
k=1,\dots,K,
\ee
while $v_k\sim\textnormal{Beta}(1,\alpha_0)$ for any $k>K$.

\subsection{Update of indicator variables $\xi_1,\dots,\xi_n$}
\label{sec:computational:update:xi}

The full conditional distribution of indicator variables $\xi_i$ is such that
\be
\Pr\left\{\xi_i=k\g \cdot\right\}
\propto
f(\bx_i\g\btheta_{\xi_i}^*)\,\mathbbm{1}(u_i<\omega_{\xi_i})
\propto
\begin{cases}
	\,\, f(\bx_i\g\btheta_i^*) & u_i<\omega_{\xi_i}, \\
	\,\, 0 & \textnormal{otherwise}.
\end{cases}
\ee

\subsection{Update of DAG and Cholesky parameters}
\label{sec:computational:update:PAS}

Under the full conditional distribution the cluster-specific parameters $\{\btheta_k^*=(\bmu_k^*,\bOmega_k^*,\D_k^*)$, $k=1,2,\dots,$\} are mutually stochastically independent. 
We can therefore update separately each component, and obtain
\ben
\label{eq:full:cond:theta}
p(\btheta_k^*\g\cdot)\propto
\begin{cases}
	\prod\limits_{i:\xi_i=k} f(\bx_i\g\btheta_k^*)p(\btheta_k^*)
	& \text{if $k \le \max\{\xi_i,i=1,\dots,n\}$},\\
	\, p(\btheta_k^*)
	& \text{otherwise}.
\end{cases}
\een
Notice that in the second line of \eqref{eq:full:cond:theta} we require to sample from the baseline over DAGs and parameters; see Section \ref{sec:computational:sampling:D} for details.
Write now $\prod_{i:\xi_i=k} f(\bx_i\g\btheta_k^*)=f(\bX^{(k)}\g\btheta_k^*)$, where $\bX^{(k)}$ is the $(n_k,q)$ matrix collecting only those observations $\bx_i$'s such that $\xi_i=k$.
Without loss of generality, consider a generic cluster and  omit for simplicity subscripts $k$ and $``^*"$ from $\btheta_k^*$ and $\bX^{(k)}$.
Update of $\btheta=(\bmu,\bOmega,\D)$
can be performed by resorting to an MCMC scheme based on a \emph{Partial Analytic Structure} (PAS) algorithm \citep{Godsill:2012}.
Consider first the re-parameterization $(\bmu,\bOmega)\mapsto(\bD,\bL,\boldsymbol{\eta})$.
The update of DAG $\D$ and parameters $(\bD,\bL,\boldsymbol{\eta})$ is then performed in two steps.

\vspace{0.2cm}
In the first step, for a given DAG $\D$, a new DAG $\widetilde{\D}$ is proposed from a suitable proposal distribution which is defined as follows.
We consider three types of operators that locally modify a DAG: insert a directed edge (InsertD $u \rightarrow v$
for short), delete a directed edge (DeleteD $u\rightarrow v$) and reverse a directed edge (ReverseD
$u \rightarrow v$). For a given $\D \in \mathcal{S}_q$, where $\mathcal{S}_q$ is the set of all DAGs on $q$ nodes, we  construct the set of valid operators $\mathcal{O}_{\D}$, that is operators whose resulting graph is a DAG.
A DAG $\widetilde{\D}$ is then called a \emph{direct successor} of $\D$ if it can be reached by applying an operator in $\mathcal{O}_{\D}$ to $\D$.
Therefore, given the current $\D$ we propose $\widetilde{\D}$ by uniformly sampling an element in $\mathcal{O}_{\D}$ and applying it to $\D$. Since there is a one-to-one correspondence between each operator and the resulting DAG, the probability of transition is $q(\widetilde{\D}\g\D)=1/|\mathcal{O}_{\D}|$, for each $\widetilde{\D}$ direct successor of $\D$.
For any two DAGs $\D$ and $\widetilde{\D}$ differing by one edge $(u,v)\in\D$,
$(u,v)\notin\widetilde{\D}$,
it can be shown that
the acceptance probability for $\widetilde{\D}$ under a PAS algorithm is given by
$\alpha_{\widetilde{\D}}=\min\{1;r_{\widetilde{\D}}\}$
with
\ben
r_{\widetilde{\D}}=
\frac{m_{}(\bX_v\g\bX_{\pa_{\widetilde{\D}}(v)}, \widetilde{\D})}
{m_{}(\bX_v\g\bX_{\pa_{\D}(v)}, \D)}
\cdot\frac{p(\widetilde{\D})}{p(\D)}
\cdot\frac{q(\D\g\widetilde{\D})}{q(\widetilde{\D}\g\D)},
\een
with
$m(\bX_v\g\bX_{\pa_{\widetilde{\D}}(v)}, \widetilde{\D})$ as in Equation \eqref{eq:marg:like:j}.

\vspace{0.2cm}
In the second step we then sample $(\bD,\bL,\boldsymbol{\eta})$
conditionally on the accepted DAG, say $\D$, from its full conditional distribution.
The latter reduces to
\be
p(\bD,\bL,\boldsymbol{\eta}\g\bX,\D)
=\prod_{j=1}^{q}
p(\bD_{jj},\bL_{\prec j \, ]},\eta_j\g\bX,\D),
\ee
where dependence on $\D$ has been made explicit. Its expression
can be recovered from the posterior of node-parameters in Section \ref{sec:posterior:distribution}.

\subsection{Update of precision parameter $\alpha_0$}

A useful property of the DP mixture model is that  $\alpha_0$ is conditionally independent of $\bX$ given $K$, parameters $\{\btheta_i\}_{i=1}^n$, and indicator variables $\{\xi_i\}_{i=1}^n$ \citep{Escobar:West:1995}. Furthermore, $\{\btheta_i\}_{i=1}^n$ are also conditionally independent of $\alpha_0$ given $K$ and the indicator variables.
Therefore, the full conditional distribution of $\alpha_0$ reduces to
$
p(\alpha_0\g K) \propto p(\alpha_0)p(K\g\alpha_0),
$
where
\be
p(K\g\alpha_0) = c_n(K) n! \alpha_0^K \frac{\Gamma(\alpha_0)}{\Gamma(\alpha_0+n)}
\ee
is the prior on $K$ implied by the DP and $c_n(K)$ is a normalizing constant not involving $\alpha_0$; see again \citet{Escobar:West:1995}.
In particular, it can be shown that the conditional posterior of $\alpha_0$ is a mixture of two Gamma densities
\ben
\label{eq:full:cond:alpha0}
\alpha_0\g\eta,K &\propto&
g \cdot \textnormal{Gamma}(c+K, d - \log\eta) +
(1-g) \cdot \textnormal{Gamma}(c+K-1, d - \log\eta),
\een
where $g/(1-g)= (c+K-1)/[n(d-\log\eta)]$ and $\eta\sim \textnormal{Beta}(\alpha_0+1,n)$.
As a consequence, at each step of the MCMC, update of $\alpha_0$ is performed by i) sampling $\eta$ conditionally on the current value of $\alpha_0$ from $\textnormal{Beta}(\alpha_0+1,n)$; ii) sampling a new value for $\alpha_0$ conditionally on $\eta$ and $K$ from \eqref{eq:full:cond:alpha0}.

\subsection{Sampling from the baseline over DAGs}
\label{sec:computational:sampling:D}

Since the enumeration of all DAGs on $q$ nodes is unfeasible in practice, direct sampling from the baseline $p(\D)$ can be achieved by adopting the following MCMC strategy.
For a given DAG $\D$ let $N(\D)$ be the set of all its direct successors, each one obtained by applying an operator in the set $\mathcal{O}_{\D}$ defined in Section \ref{sec:computational:update:PAS}.
We first uniformly sample a DAG $\widetilde{\D}$ from $N(\D)$ that is with probability $q(\widetilde{\D}\g\D)=1/|N(\D)|$, for each $\widetilde{\D}\in N(\D)$. Hence, we move to $\widetilde{\D}$ with probability
\be
\alpha_{\widetilde{\D}}=
\min
\left\{
1; \frac{p(\widetilde{\D})}{p(\D)}\cdot \frac{q(\D\g\widetilde{\D})}{q(\widetilde{\D}\g\D)}
\right\}.
\ee
Notice the similarity with the acceptance ratio in (13); clearly, the difference is that here we are sampling from the \emph{prior} over $\mathcal{S}_q$ (the set of all DAGs on $q$ nodes) and therefore no data are involved. \black
Also, to compute $\alpha_{\widetilde{\D}}$ we only need to evaluate the \emph{ratio} of the priors
$p(\widetilde{\D})/p(\D)=r$, which does not require the computation of normalizing constants over the space of DAGs and is directly available from Equation (16) in our paper.
Moreover, the ratio of the two proposal reduces to
$q(\D\g\widetilde{\D})/q(\widetilde{\D}\g\D)=|\mathcal{O}_{\D}|/|\mathcal{O}_{\widetilde{\D}}|$
which instead requires the enumeration of all the direct successors of $\D$ and $\widetilde{\D}$. While this is feasible with a relatively small computational cost, it was observed empirically that the approximation
$q(\D\g\widetilde{\D})/q(\widetilde{\D}\g\D)\approx 1$
does not produce a relevant loss in terms of accuracy.

\section{Analysis of AML data}
\label{sec:supplementary:application:AML}

\subsection{Diagnostics of convergence}

To assess the convergence of our MCMC algorithm on the AML data we ran two independent chains.
Under both replicates we set prior hyperparameters as in Section 6 of our paper and fix the number of MCMC iterations to $S=120000$, which includes a burn-in period of $20000$ runs.

\begin{figure}
	\begin{center}
		\begin{tabular}{cc}
			\includegraphics[scale=0.44]{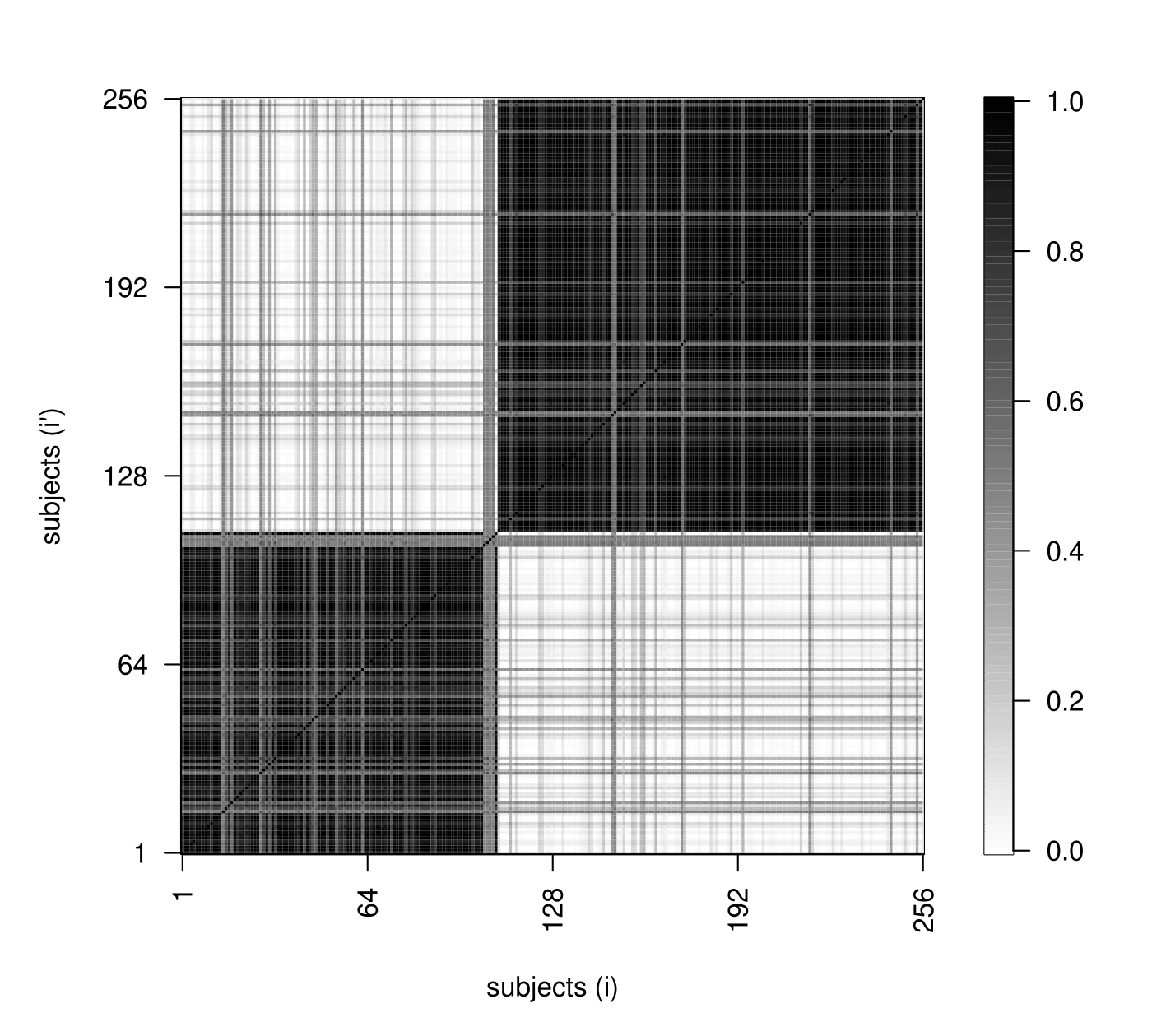}
			\includegraphics[scale=0.44]{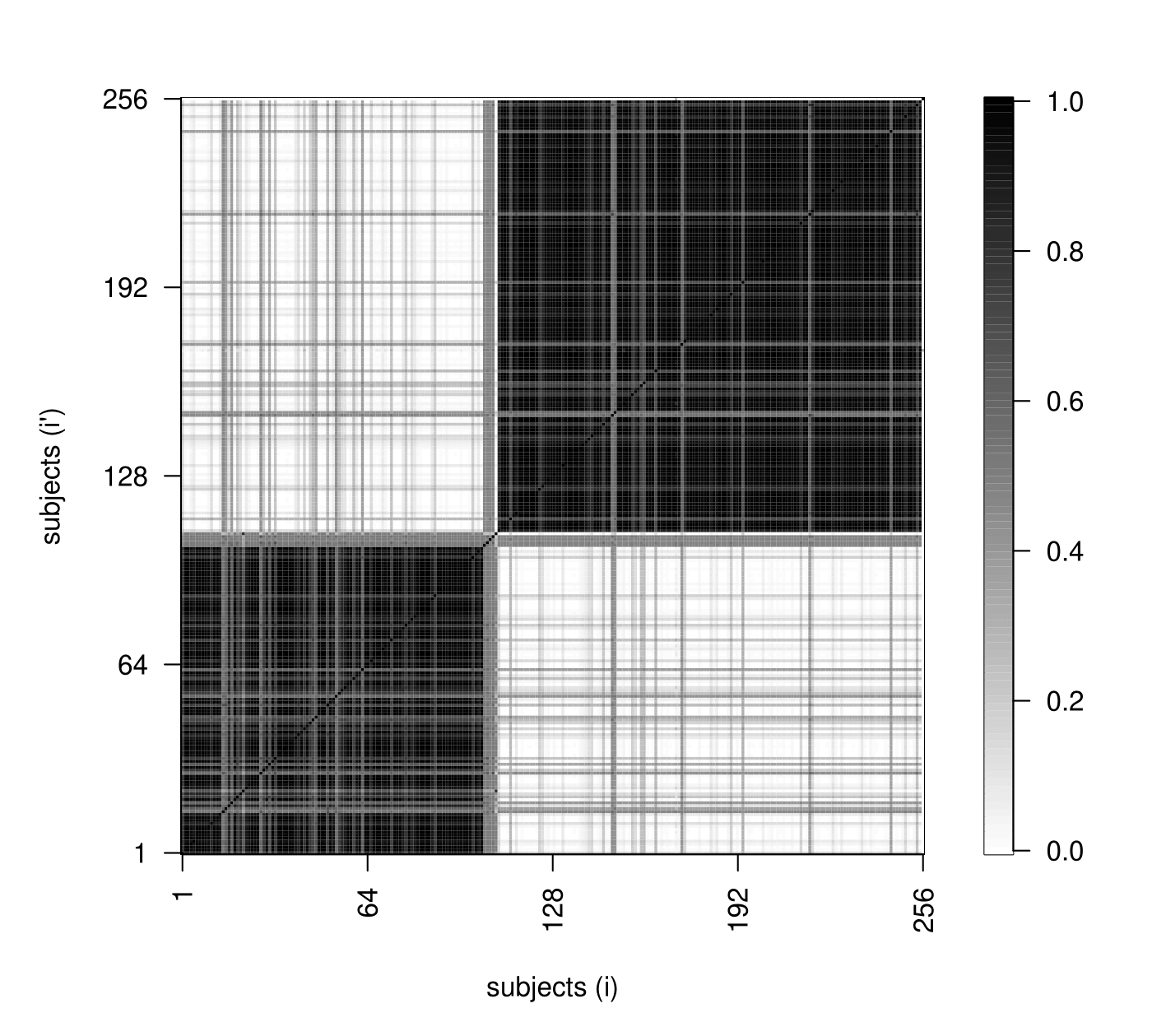}
		\end{tabular}
		\caption{\small AML data. Posterior similarity matrices obtained from two independent MCMC chains. Subjects arranged by cluster membership.}
		\label{fig:application:suppl:diag:1}
	\end{center}
\end{figure}

For each of the two MCMC outputs we first produce an $(n,n)$ similarity matrix based on the approximate posterior probabilities in Equation (17) of our paper, computed for each pair of subjects $(i,i'), i\ne i'$.
The two resulting matrices are represented as heat-maps in Figure \ref{fig:application:suppl:diag:1}.
By inspection, it appears that the agreement between the two outputs is highly satisfactory.
Also, the two cluster structures, obtained by fixing a threshold for inclusion of $0.5$, coincide.

\begin{figure}
	\begin{center}
		\begin{tabular}{c}
			\includegraphics[scale=0.51]{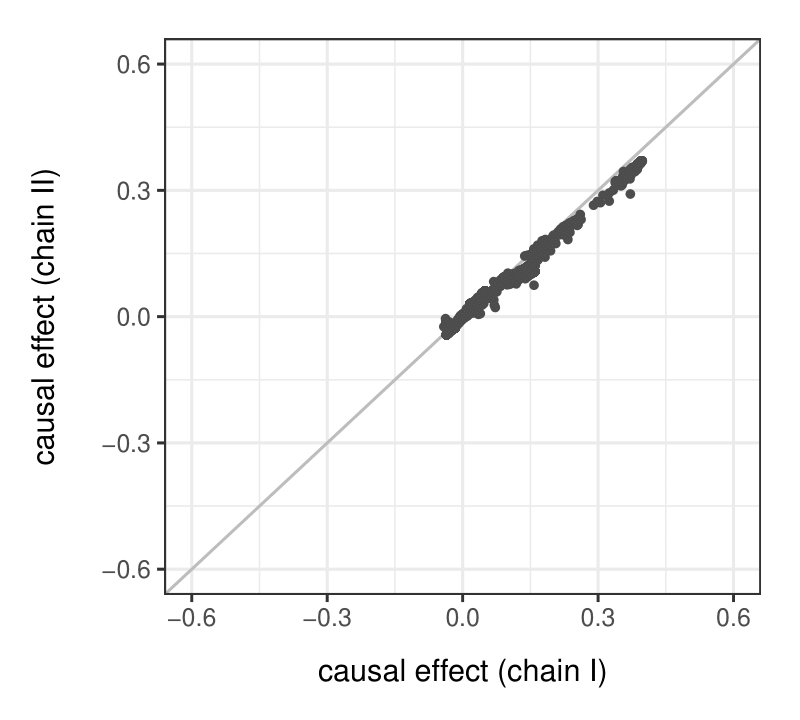}
			\includegraphics[scale=0.51]{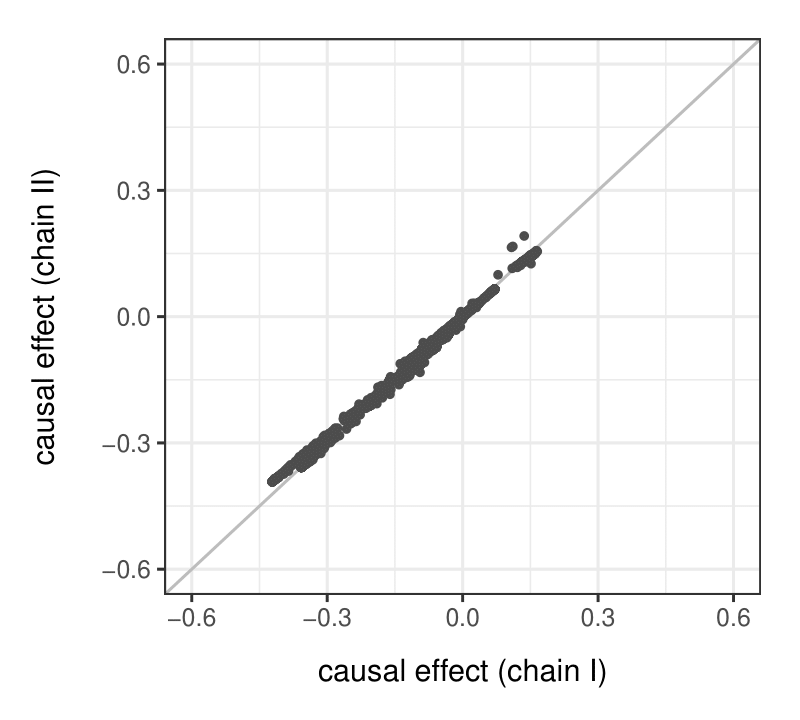}
			\includegraphics[scale=0.51]{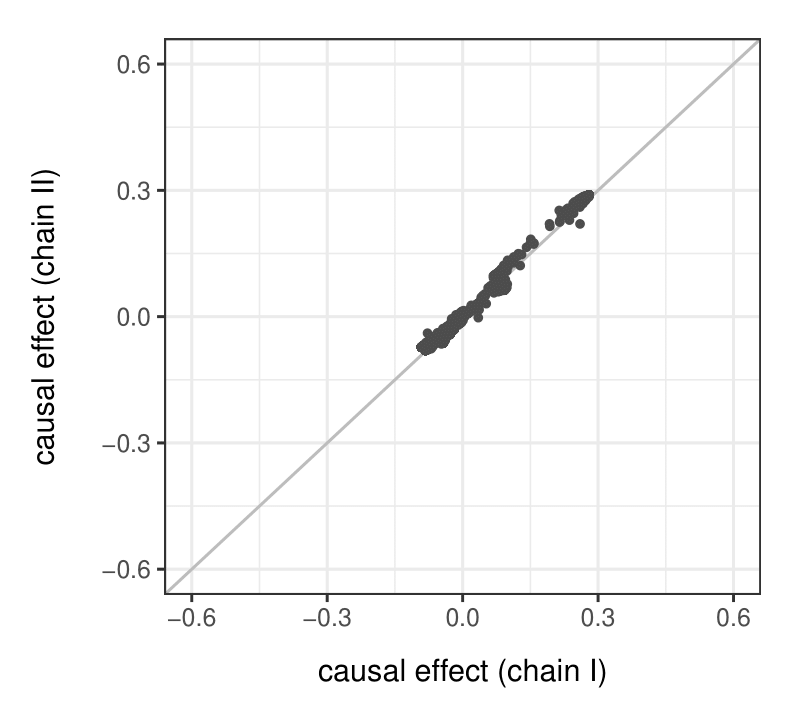}
		\end{tabular}
		\caption{\small AML data. Scatter plots of the BMA causal-effect estimates for response nodes AKT, AKT.p308, AKT.p473 (from left to right) obtained from two independent MCMC chains.}
		\label{fig:application:suppl:diag:2}
	\end{center}
\end{figure}

Under each MCMC output, we then produce the collection of subject-specific BMA causal effect estimates.
Figure \ref{fig:application:suppl:diag:2} shows the scatter plots of causal effect estimates for response nodes AKT, AKT.p308 and AKT.p473 (computed across all possible intervened nodes) obtained from the two MCMC chains. One can see that the agreement between the two outputs is highly satisfactory, since points lie close to the 45-degree reference line.

\subsection{Comparison with alternative clustering strategies}

Our method accounts for population heterogeneity as it naturally induces a clustering of the observations in an unsupervised way.
Alternative approaches can make use of covariates, whenever these are available, to first group observations/patients into homogeneous clusters.
Next, inference on cluster-specific DAG structures and causal effects can be implemented for each group separately as in the \textit{Oracle} version of our method introduced for comparison purposes in the simulation study of Section 5.

In the dataset here considered, patients are classified into $11$ subtypes by the French-American-British (FAB) system, which relies mainly upon morphologic features; see also Section 6 of our paper.
We first investigate potential agreement 
 between the clustering structure estimated by our DP mixture method and the groups defined by the AML subtypes.

\begin{figure}
	\begin{center}
		\begin{tabular}{c}
			\includegraphics[scale=0.65]{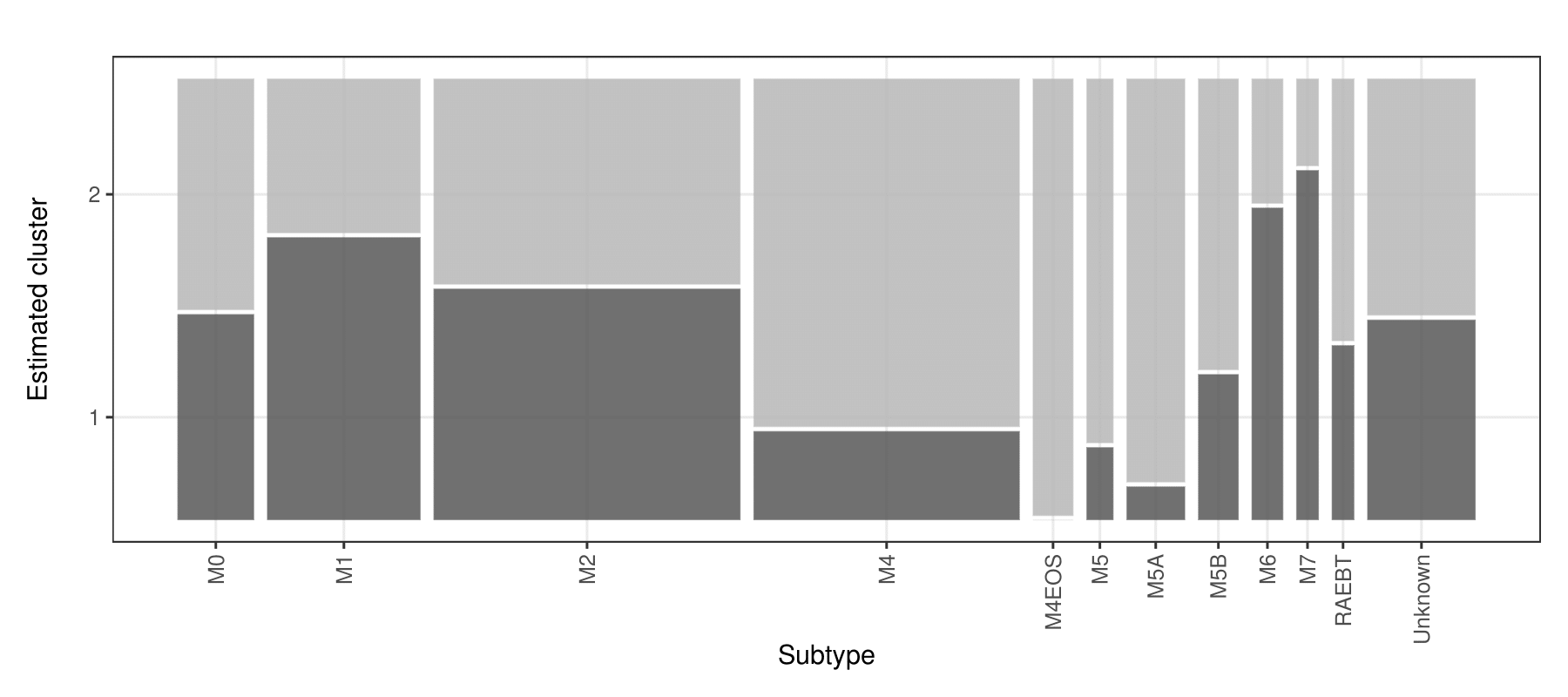}
		\end{tabular}
		\caption{\small AML data. Mosaic plot with the proportion of patients assigned to estimated clusters 1 and 2 (dark and light grey respectively) for each AML subtype.}
		\label{fig:application:cfr:subtypes}
	\end{center}
\end{figure}

Table \ref{tab:application:cfr:subtypes} reports, for each subtype, the percentage of patients that are assigned to clusters 1 and 2 by our method.
To better appreciate this comparison, a graphical representation of the output is provided in the mosaic plot of Figure \ref{fig:application:cfr:subtypes}. Here in particular, the width of each bar representing one of the subtypes is proportional to the corresponding group (subtype) sample size. The dark grey and light grey areas represent the proportion of patients belonging to clusters 1 and 2 respectively.
It appears that the allocation to Cluster 1 or 2 is quite variable among the subtypes. In particular for groups M0 and M2  there is almost an even splitting, while the splitting is much more unbalanced for   M4EOS, M5, M5A and to a lesser extent for M4.

\begin{table}[ht]
	\centering
	\footnotesize
	\begin{tabular}{ccccccccccccc}
		\hline
		Subtype & M0 & M1 & M2 & M4 & M4EOS & M5 & M5A & M5B & M6 & M7 & RAEBT & Unknown \\
		\hline
		size & 17 & 34 & 68 & 59 & 9 & 6 & 13 & 9 & 7 & 5 & 5 & 24 \\
		\hline
		\% Cluster 1 & 47\% & 65\% & 53\% & 20\% & 0\% & 17\% & 8\% & 33\% & 71\% & 80\% & 40\% & 46\% \\
		\% Cluster 2 & 53\% & 35\% & 47\% & 80\% & 100\% & 83\% & 92\% & 67\% & 29\% & 20\% & 60\% & 54\% \\
		\hline
	\end{tabular}
\caption{\small AML data. AML subtypes defined according to the French-American-British (FAB) system and corresponding number of subjects (size) in the dataset; percentage of patients within each subtype assigned to estimated clusters 1 and 2 by method DP mixture.}
\label{tab:application:cfr:subtypes}
\end{table}

We now implement our method in its \textit{Oracle} version for each group/subtype independently to infer group-specific causal effects.
Clearly, 
 the battery of estimated causal effects will be equal  
for all patients assigned to the same subtype.
In the following, we focus on the $n_{M4} = 59$ patients classified as subtype M4 and compare the corresponding subject-specific causal effects obtained from our DP mixture model with the M4 cluster-specific causal effects (\textit{Oracle M4} in the following).

\begin{figure}
	\begin{center}
		\begin{tabular}{c}
			\includegraphics[scale=0.40]{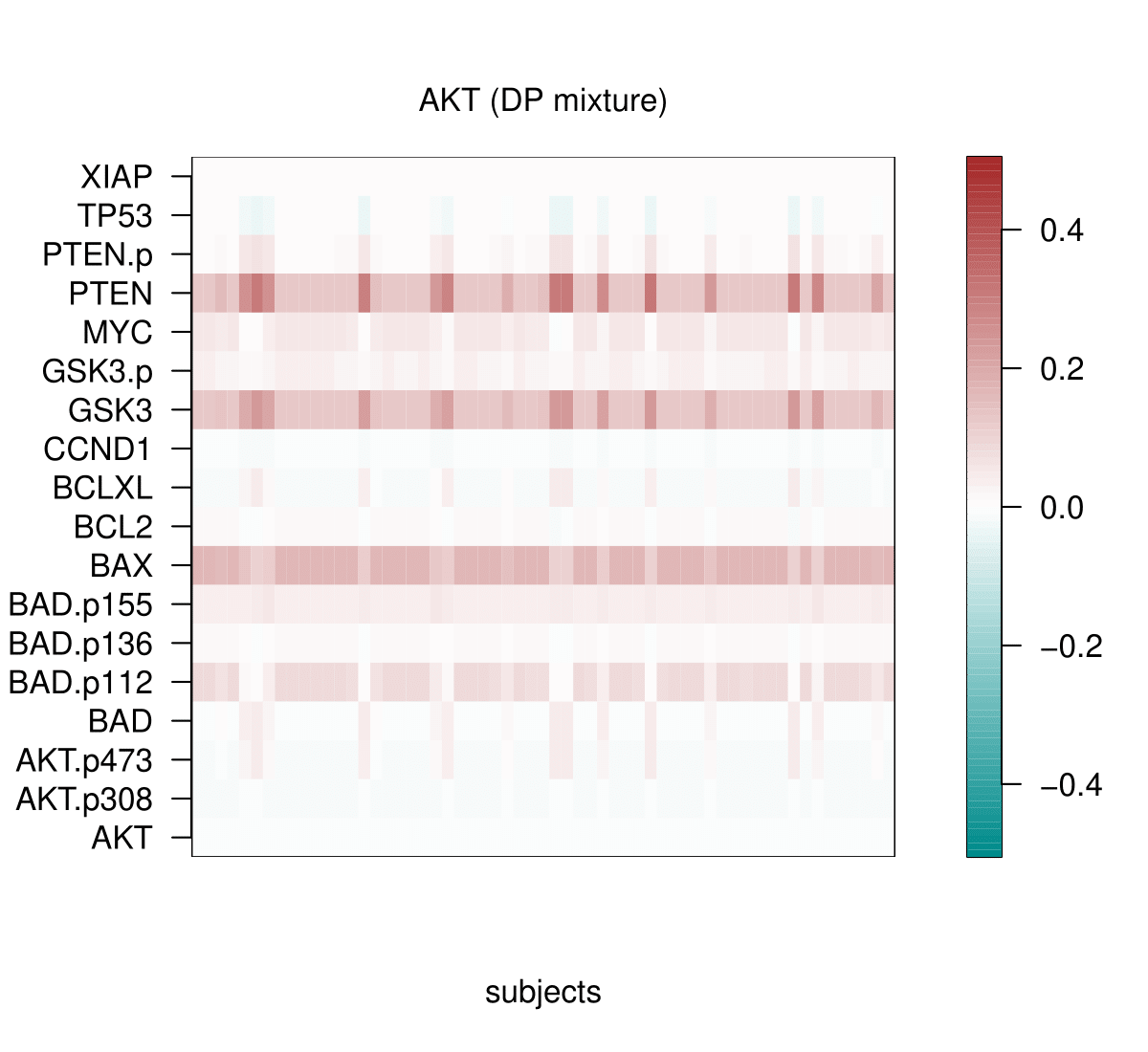}
			\includegraphics[scale=0.40]{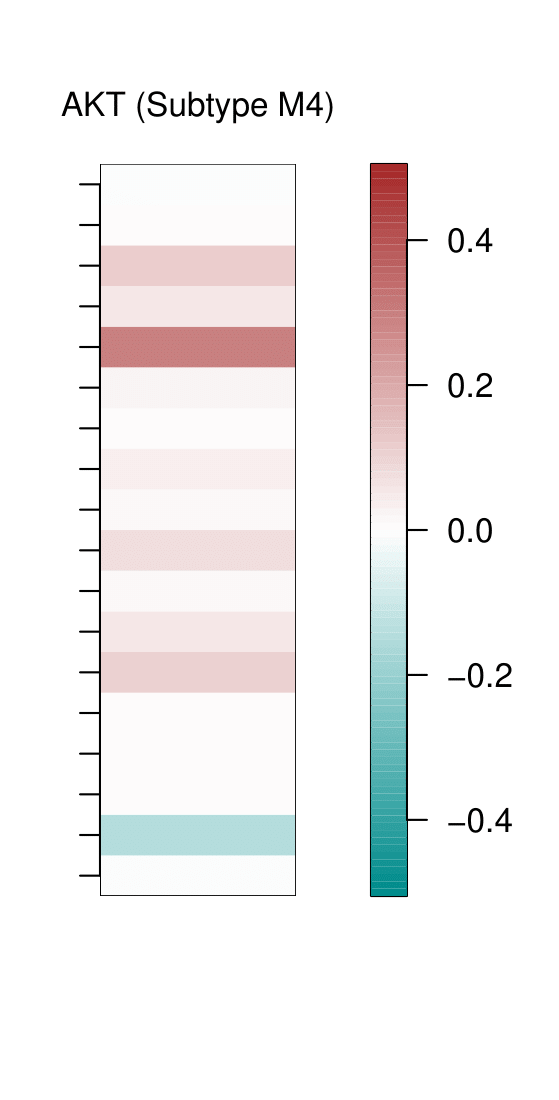} \\
			\includegraphics[scale=0.40]{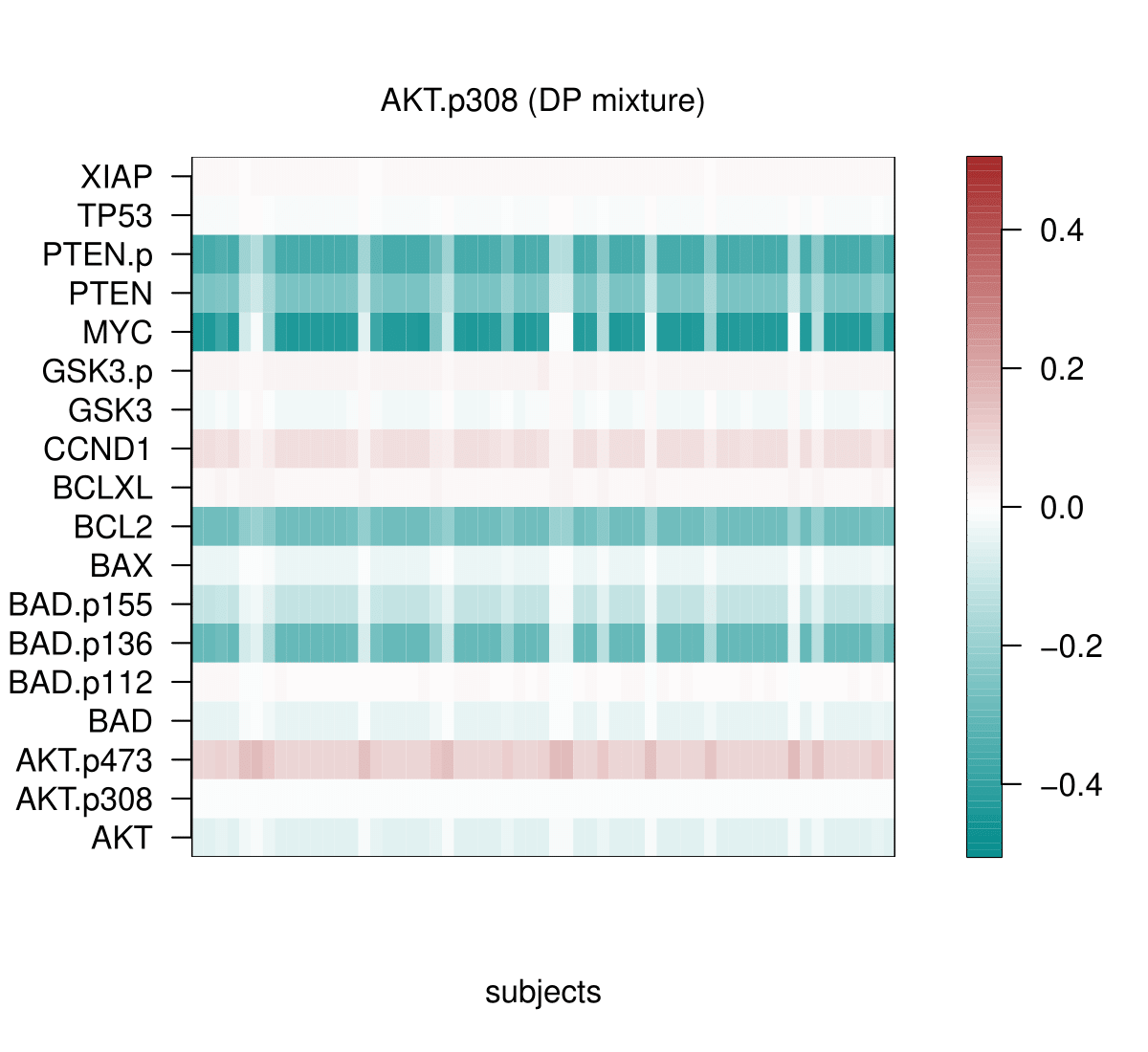}
			\includegraphics[scale=0.40]{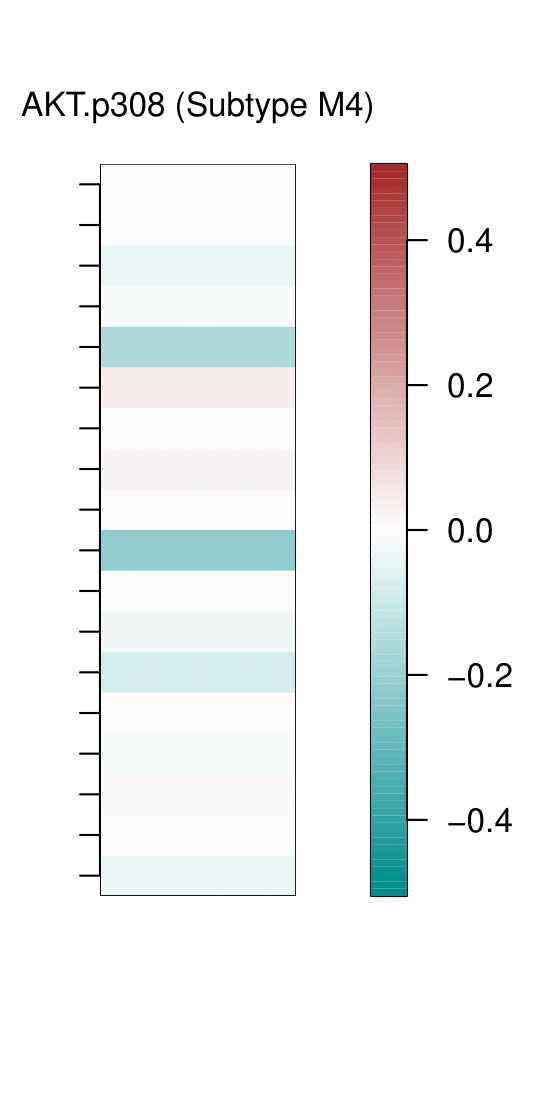} \\
			\includegraphics[scale=0.40]{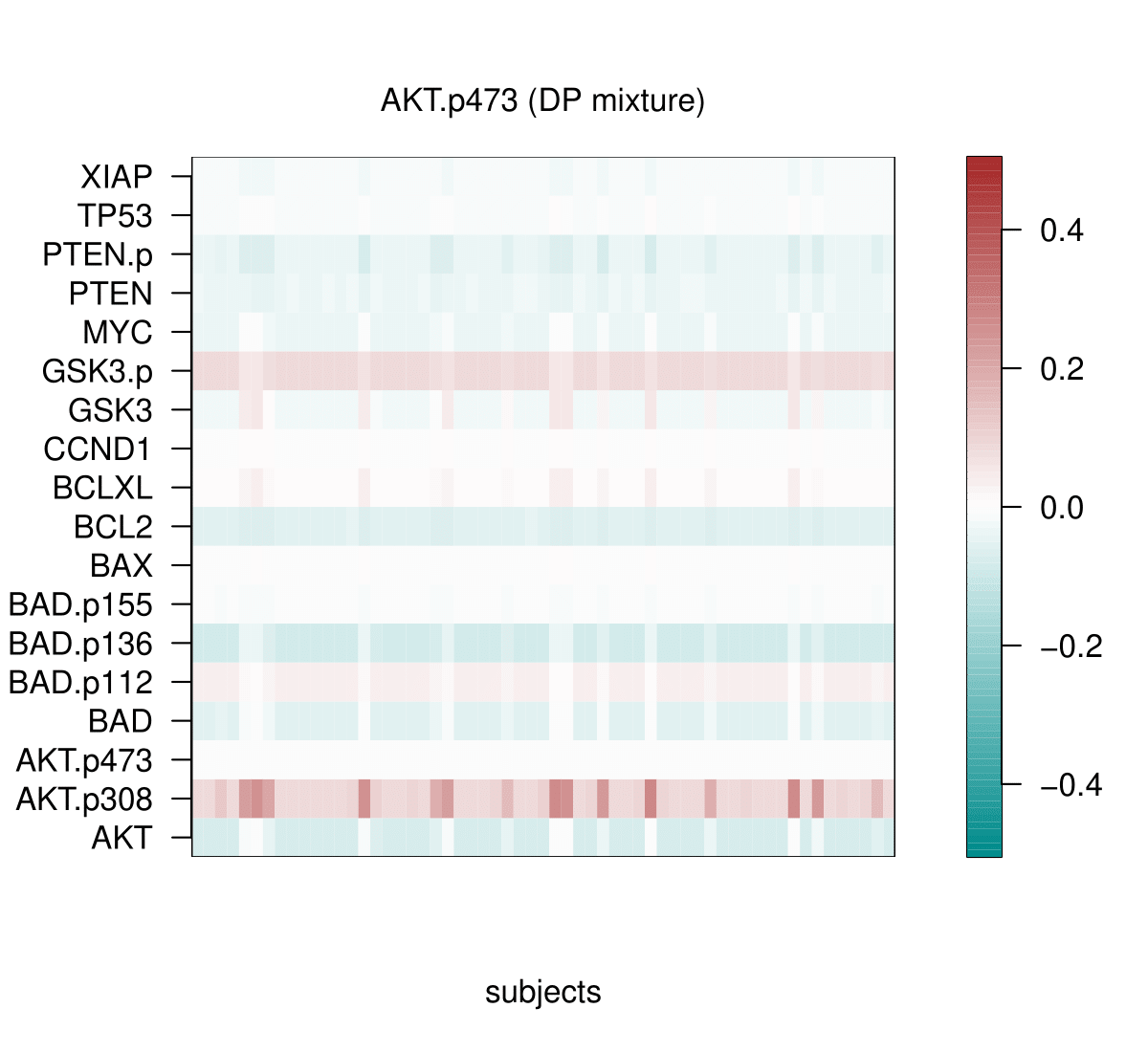}
			\includegraphics[scale=0.40]{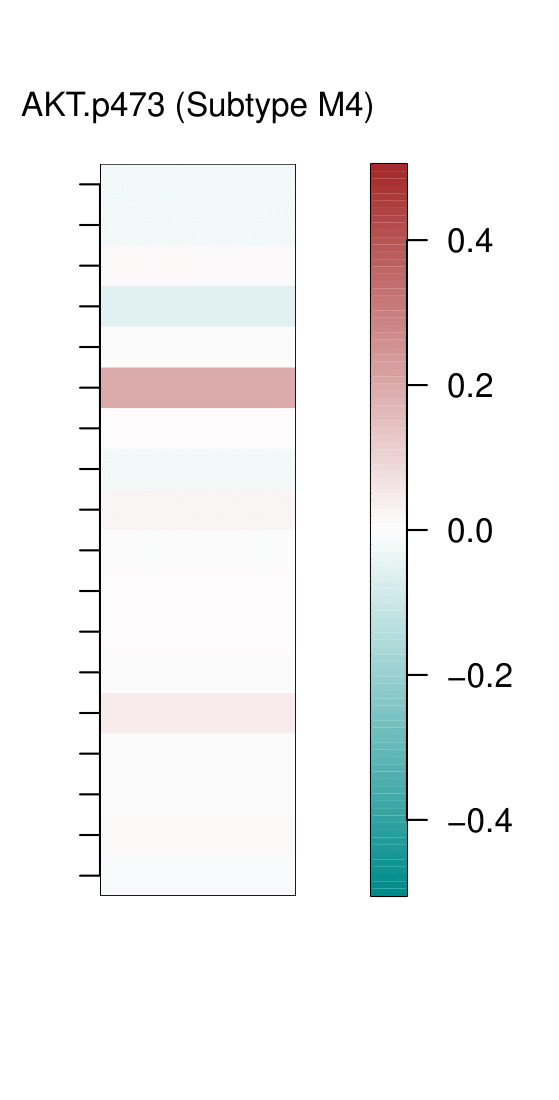} \\
		\end{tabular}
		\caption{\small AML data. Heat maps of causal effects on responses AKT, AKT.p308, AKT.p473, following an intervention on one target protein among the 18 in the network (AKT, ..., XIAP) for subjects assigned to subtype M4. Left-side heat maps report the DP mixture causal effect estimates; right-side heat maps refer to the Oracle M4 causal effect estimates.}
		\label{fig:application:causal:effects:M4}
	\end{center}
\end{figure}

Results are summarized in the heat maps of Figure \ref{fig:application:causal:effects:M4} where each row-block  refers to one of the three response variables among AKT, AKT.p308 and AKT.p473.
In particular, left panels report the BMA subject-specific causal effect estimates for the  M4-patients resulting  from our DP mixture model.
It appears that most  subjects exhibit very similar  causal effects; indeed these individuals represent the 80\% M4-patients assigned to the estimated cluster 2; see also Table \ref{tab:application:cfr:subtypes}.
More interestingly however, there are substantial differences
with respect to the collection of M4 cluster-specific causal effect estimates reported in the right side heat maps.
This result suggests that
methods which fully rely on a pre-defined clustering may produce inadequate or misleading estimates of causal effects.
In addition, we emphasize that subtypes with a small number of subjects can produce unreliable estimates because the data provide a weak source of information.

\black

\bibliographystyle{Chicago}
\bibliography{biblio}
